\documentclass[11pt]{article}
\usepackage{amsmath, amsfonts, amssymb, amsthm, newlfont, graphicx}

\numberwithin{equation}{section}

\newcommand{\pa}{\partial}

\newcommand{\be}{\begin{equation}}
\newcommand{\ee}{\end{equation}}

\newcommand{\bt}{\mathbf{t}}

\begin{document}

\title{Quantum Hasimoto transformation and nonlinear waves on a superfluid vortex filament under the quantum local induction approximation}

\author{Robert A. Van Gorder$^*$ \\  
\small Mathematical Institute, University of Oxford\\
\small Andrew Wiles Building, Radcliffe Observatory Quarter, Woodstock Road, Oxford, OX2 6GG, United Kingdom\\
\small $^*$Corresponding author. Email: Robert.VanGorder@maths.ox.ac.uk}        
\date{\today}       
\maketitle

\begin{abstract}
The Hasimoto transformation between the classical LIA (local induction approximation, a model approximating the motion of a thin vortex filament) and the nonlinear Schr\"odinger equation (NLS) has proven very useful in the past, since it allows one to construct new solutions to the LIA once a solution to the NLS is known. In the present paper, the quantum form of the LIA (which includes mutual friction effects) is put into correspondence with a type of complex nonlinear dispersive partial differential equation (PDE) with cubic nonlinearity (similar in form to a Ginsburg-Landau equation, with additional nonlinear terms). Transforming the quantum LIA in such a way enables one to obtain quantum vortex filament solutions once solutions to this dispersive PDE are known. From our quantum Hasimoto transformation, we determine the form and behavior of Stokes waves and a standing 1-soliton solution under normal and binormal friction effects. The soliton solution on a quantum vortex filament is a natural generalization of the classical 1-soliton solution constructed mathematically by Hasimoto (which motivated subsequent real-world experiments). We also find that self-similar solutions are possible under this quantum Hasimoto transform, and these solutions demonstrate that even in the presence of dissipative effects, there can be strong local increases in curvature or torsion even though asymptotically the solutions are smoothed through dissipation. More general traveling wave solutions propagating along the quantum filament with fixed wave speed are also studied, with such solutions holding the 1-soliton as a special case. The quantum Hasimoto transformation is useful when normal fluid velocity is relatively weak, so for the case where the normal fluid velocity is dominant we resort to other approaches. We consider the dynamics of the tangent vector to the vortex filament directly from the quantum LIA, and this approach, while less elegant than the quantum Hasimoto transformation, enables us to study waves primarily driven by the normal fluid velocity. We exhibit a number of solutions that exist only in the presence of the normal fluid velocity and mutual friction terms (which would therefore not exist in the limit taken to obtain the classical LIA, decaying into line filaments under such a limit), examples of which include normal fluid driven helices, stationary and propagating topological solitons, and a vortex ring whose radius varies inversely with the normal fluid magnitude. We can also view the dynamics of a traveling wave solution for the tangent vector in terms of a dynamical system on the unit sphere, and this formulation gives us another framework by which to obtain solutions. We show that, while chaos may not be impossible under the quantum LIA, it should not be expected to arise from traveling waves along quantum vortex filaments under the quantum LIA formulation. We also conjecture on the possibility of chaos in related systems, and on the existence more complicated solitons such as breathers.
\end{abstract}

\noindent \textit{Keywords}:

\section{Introduction}
Unlike in a classical fluid, the motion of vortex filaments in a superfluid such as $^4$He is modeled taking into account mutual friction, such as that discussed in the HVBK model (Hall and Vinen 1956a,b; Bekarevich and Khalatnikov, 1961). A number of fundamental results were provided by Vinen (1957a,b,c, 1958). Mutual friction also plays a role in the dynamics of vortex filaments in $^3$He (Kopnin and Manninen, 1991; Bevan et al, 1995). The self-induced motion of a single quantized vortex filament is governed by the Biot-Savart integral. Nonlocal dynamics under the Biot-Savart integral are often approximated by the local induction approximation (LIA), provided that such an approximation is reasonable for the specific physical scenario being described. At finite temperatures, a quantized vortex is also influenced by mutual friction. Replacing the non-local term with the LIA and accounting for mutual friction in the normal and binormal directions, Schwarz (1985) obtained a kind of quantum LIA (qLIA) which accounts for mutual friction and the interaction with a normal fluid. This model is given in non-dimensional form by 
\be \label{fil}
\textbf{v} =  \kappa \mathbf{t} \times \mathbf{n} + \alpha \mathbf{t} \times (\mathbf{U} -  \kappa \mathbf{t}\times\mathbf{n}) - \alpha^\prime \mathbf{t} \times (\mathbf{t} \times (\mathbf{U} -  \kappa \mathbf{t}\times\mathbf{n}))\,.
\ee
Here $\mathbf{U}$ is the dimensionless normal fluid velocity, $\mathbf{t}$ and $\mathbf{n}$ are the unit tangent and unit normal vectors to the vortex filament, $\kappa$ is the curvature, $\alpha$ and $\alpha^\prime$ are dimensionless friction coefficients which are small (except near the $\lambda$-point; for reference, the $\lambda$-point is the temperature ($\approx 2.17\text{K}$, at atmospheric pressure) below which normal fluid Helium transitions to superfluid Helium (Landau and Lifshitz, 1959)). In the limit $(\alpha ,\alpha')\rightarrow (0,0)$, we recover the classical Da Rios equations for the motion of a vortex filament in a classical fluid (Da Rios, 1906; Arms and Hama, 1995; Ricca, 1991). 

A number of studies exist on the solutions to the quantum LIA. In the $\alpha,\alpha' \rightarrow 0$ limit, these solutions should collapse into solutions of the classical LIA. One highly important class of solutions to the classical LIA would be the 1-soliton solution found by Hasimoto (1972), by way of what is now referred to as the Hasimoto transformation, which puts the classical LIA into correspondence with the cubic NLS. While a number of solutions to the quantum LIA have been studied either numerically or analytically, the Hasimoto 1-soliton have never been extended to the quantum LIA. The purpose of this paper is to fill this important gap. Applying a method analogous to that of Hasimoto, we are able to put the quantum LIA \eqref{fil} into correspondence with a type of complex Ginzburg-Landau equation (GLE), a natural complex-coefficient generalization of NLS. From this, we study Stokes waves, 1-solitons, and other traveling wave solutions. Each of these solutions generalizes known results for the classical LIA. We also conjecture that there is the possibility of chaos in such systems. Similarity solutions are also demonstrated.

The obtained Hasimoto type transform for the quantum LIA (including mutual friction terms), which puts the quantum LIA into correspondence with a complex GLE, is a completely natural generalization of the classical LIA to cubic NLS map pioneered by Hasimoto, which in its own right has motivated both experimental and theoretical work over the last 40 years. As one example, Hopfinger and Browand (1982) studied turbulent flow in a rotating container, and determined that the theoretical solutions predicted by Hasimoto (1972) do actually occur along vortex filaments in practical experiments. It therefore makes sense that solitons, such as the one obtained here from the quantum LIA, should occur along quantum filaments. While much interest has been directed at the study of quantum vortex filament dynamics, studies on solitons and other waves on such filaments often consider the classical case (the zero-temperature limit), and neglect any mutual friction parameters. However, in this paper, we demonstrate that it is possible to obtain solitons (and other waves) on a vortex filament in the more complicated quantum model, and we discuss the qualitative effects of the additional parameters inherent in this model on such solutions. Since the complex GLE has been shown to admit a variety of dynamics in the last 20 years, by accounting for the quantum LIA we permit a far wider variety of solutions that would be possible from the classical LIA results in the literature. This opens the literature in this area up to far more interesting and realistic vortex filament dynamics, which in turn can be of use as motivation or comparison for those conducting experiments on waves along vortex filaments. 

The paper roughly falls into three parts, and is organized as follows. The first part of this paper involves deriving the quantum Hasimoto transformation. To this end, in Section 2, we give a mathematical formulation for the quantum LIA in terms of a kind of complex Ginsburg-Landau equation (GLE) with additional damping terms. To obtain this complex scalar partial differential equation, we construct a quantum analogue of the Hasimoto transformation. We should note that in order to obtain the quantum Hasimoto transformation, it is necessary to assume that normal fluid velocity effects are small enough to be negligible. This is not simply for convenience; if this assumption is not made, the quantum Hasimoto transformation does not exist. While somewhat undesirable, such solutions still maintain physical relevance (Almari, Youd, Barenghi, 2008; Araki, Tsubota, Nemirovshii, 2002; Vinen, 2001; Van Gorder, 2014a), particularly in the low-temperature limit.

In the second part of this paper, we obtain various vortex filament solutions using the PDE derived through the quantum Hasimoto transformation. In Section 3, we discuss the quantum analogue of Stokes waves along a vortex filament, and we demonstrate that such waves have an algebraic rate of decay in the quantum case. We then consider the existence of a quantum generalization of the Hasimoto 1-soliton solution in Section 4. We find a standing ``soliton" which maintains its form to order $\alpha$, with dissipative effects confined to order $\alpha^2$ or lower terms. This means that such a standing wave should, in principle, exist for sufficiently long timescales to be detected experimentally, before dissipative effects collapse the solution. We next discuss self-similar solutions under the Hasimoto transformation in Section 5. The structure of these solutions agree with what is known for other formulations of the quantum LIA. Interestingly, while the quantum Hasimoto transformation includes mutual friction terms which tend to damp the solutions as time progresses, these self-similar solutions demonstrate that at least locally there can be increases in curvature and torsion over time, before the damping finally dominates. Traveling wave solutions, which propagate along the quantum vortex filament, are discussed in Section 6. These would hold the standing soliton as a special case (when wave speed tends to zero). These solutions can have infinitely many local curvature maxima (as opposed to the unique curvature maxima exhibited by the 1-soliton), depending on the model parameters taken. As we should expect, it is shown that mutual friction effects have a damping effect on these waves (compared to the corresponding traveling waves for the classical LIA which has no such friction parameters).

In the final part of this paper, we consider some solutions which cannot be found using the derived quantum Hasomoto transformation, as they depend on the normal fluid velocity in a structurally fundamental way. Such solutions can therefore only be obtained only when we include the effects of the normal fluid. In Section 7, we discuss a potential formulation for the tangent vector to the filament, and we employ this formulation to find various waves generated by the normal fluid velocity. A scalar potential formulation for the time evolution of the tangent vector to the vortex filament under the quantum LIA was considered in Van Gorder (2014a), although the formulation we consider here is derived in a more specific way for the traveling waves. This permits us to study helical structures driven by the normal fluid velocity and some stationary structures as very special cases. This formulation comes along with a consistency constraint which must be satisfied (essentially, this constraint is due to the approximation of a three-dimensional problem using a scalar PDE), and we give an example of this condition for the helical model. In Section 8, we consider a direct approach using the tangent vector to the filament (with no potential function employed). Treating the tangent as a unit vector, we essentially are able to transfer the traveling wave problem into a dynamical system on the unit sphere. An exact stationary solution is obtained; while generated by the normal fluid velocity, this solution is eternal (in that it persists in time without change of form). In Section 9, we consider solutions propagating due to the normal fluid flow, which are obtained directly from the vector form of the quantum LIA under the assumption of a position vector parameterized in a way that depends on laboratory coordinates. As there are no simplifications taken, these solutions are governed by a more complicated dynamical system than other solutions discussed in this paper. We demonstrate that a wide class of eternal solutions exist which move with the normal fluid velocity while maintaining their form for all time. One example of such a solution is a vortex ring oriented orthogonal to, and moving with, the normal fluid velocity. This ring tightens as the magnitude of the normal fluid velocity increases. A planar topological soliton is also constructed which demonstrates this type of motion. Interesting, there is a degeneracy in the full quantum LIA under the traveling wave assumption which implies that the full dynamics are either one or two dimensional, but not fully three dimensional. This, in turn, precludes the existence of chaotic dynamics for traveling wave reductions of the quantum LIA.

A discussion and brief concluding remarks are finally given in Section 10. Areas of interest for future work, such as constructing breather solutions or finding chaos in the quantum form of the LIA, are mentioned in Section 10. While chaos is not to be observed in traveling waves, there may be other structures which could allow for chaotic dynamics to emerge.

\section{Mathematical formulation for the quantum Hasimoto transformation}
Differentiating with respect to the arclength variable $s$, and performing several vector manipulations, we have that the quantum LIA \eqref{fil} becomes
\be
\dot{\bt} = \frac{\pa }{\pa s}\left\lbrace (1-\alpha' |\bt|^2)\bt\times\bt_s - \alpha [(\bt \cdot \bt_s)\bt - |\bt|^2\bt_s]   \right\rbrace
+ \frac{\pa }{\pa s}\left\lbrace \alpha \bt\times \mathbf{U} -\alpha' \bt\times(\bt \times \mathbf{U}) \right\rbrace\,.
\ee
Taking $\bt$ to be a unit vector, the equation simplifies slightly to
\be\label{unitLIA}
\dot{\bt} = \frac{\pa }{\pa s}\left\lbrace (1-\alpha')\bt\times\bt_s + \alpha \bt_s + \alpha \bt \times \mathbf{U} - \alpha' (\bt \cdot \mathbf{U})\bt   \right\rbrace\,.
\ee
This puts the quantum LIA \eqref{fil} into the form of a vector conservation law. 

Similar results to that which we shall provide in this section were recently attempted in the case of $\mathbf{U}\neq \mathbf{0}$ (Shivamoggi, 2013), however the resulting system was not solved and only the limiting reduction to $\alpha,\alpha' =0$ (i.e., the classical case) was given. Some qualitative observations were also given at lowest order. In what follows, we shall assume that $\alpha |\mathbf{U}|=O(\alpha^2)$, i.e. that the magnitude of $\mathbf{U}$ is sufficiently small. Many studies on quantum vortex dynamics have taken the normal fluid velocity to zero (Almari, Youd, Barenghi, 2008; Araki, Tsubota, Nemirovshii, 2002; Vinen, 2001; Van Gorder, 2014a), and this is effectively the same thing as setting $\mathbf{U}=\mathbf{0}$. The physical applicability of such a scenario is limited to the very low temperature regime where the mutual friction coefficients are very small, which permits us to neglect perturbations of $O(\alpha^2)$. 

Let $\mathbf{b}$ denote the binormal vector, and take $\kappa$ and $\tau$ to be the curvature and torsion, respectively. Recall that $\bt_s =\kappa\mathbf{n}$, $\mathbf{n}_s=-\kappa\bt +\tau\mathbf{b}$, $\mathbf{b}_s=-\tau\mathbf{n}$. After setting $\mathbf{U}=\mathbf{0}$ in \eqref{unitLIA}, we can write \eqref{unitLIA} as 
\be \label{simplified}
\dot{\bt}=(1-\alpha')(\kappa\mathbf{b})_s+\alpha\bt_{ss}\,.
\ee
Let us define the function
\be \label{defn}
\psi(s,t) = \kappa(s,t)\exp\left(i\int_0^s\tau(\hat{s},t)d\hat{s}\right)
\ee
and also a new vector-valued function by $\mathbf{m}=(\mathbf{n}+i\mathbf{b})\exp(i\int_0^s\tau(\hat{s},t)d\hat{s})$. Note that $\mathbf{m}_s=-\psi\bt$ and $\bt_s = \frac{1}{2}(\psi^*\textbf{m}+\psi\mathbf{m}^*)$, where $^*$ denotes complex conjugation. Additionally, $(\kappa\mathbf{b})_s=\frac{i}{2}(\psi_s\mathbf{m}^*-\psi_s^*\mathbf{m})$. The quantum LIA \eqref{unitLIA} therefore takes the form
\be 
\dot{\bt} = \frac{i}{2}(1-\alpha')(\psi_s\mathbf{m}^*-\psi_s^*\mathbf{m})+\frac{\alpha}{2}(\psi^*\textbf{m}+\psi\mathbf{m}^*)_s\,.
\ee

We seek to derive an equation for $\psi$ in analogy to that which was obtained by Hasimoto in the case of a classical fluid (i.e., $\alpha=\alpha'=0$). On the one hand, note that
\be \begin{aligned}\label{dot1}
\dot{\mathbf{m}}_s & = -\dot{\psi}\bt -\psi\dot{\bt}\\
& = -(\dot{\psi} +\alpha|\psi|^2\psi)\bt + \frac{i(1-\alpha')-\alpha}{2}\psi\psi_s^*\mathbf{m}  - \frac{i(1-\alpha')+\alpha}{2}\psi\psi_s\mathbf{m}^*\,.
\end{aligned}\ee
On the other hand, assume that we have a representation for $\dot{\mathbf{m}}$ of the form
\be 
\dot{\mathbf{m}} = a \mathbf{m} + b \mathbf{m}^* + c \bt \,.
\ee
First, observe that
\be 
a + a^* = \frac{1}{2}(\dot{\mathbf{m}}\cdot \mathbf{m}^* + \dot{\mathbf{m}}^*\cdot \mathbf{m}) = \frac{1}{2}\frac{\pa}{\pa t}(\mathbf{m}\cdot\mathbf{m}^*) =0\,,
\ee
therefore $a$ must take the form $a=i\phi(s,t)$, for some real-valued function $\phi$. By a similar process, $b\equiv 0$. We should also find that
\be 
c = -\mathbf{m}\cdot \dot{\bt} = -(i(1-\alpha')+\alpha)\psi_s\,.
\ee
Therefore, we have the representation 
\be 
\dot{\mathbf{m}} = i\phi(s,t) \mathbf{m} - (i(1-\alpha')+\alpha)\psi_s(s,t) \bt\,.
\ee
Differentiation of this representation with respect to arclength gives
\be \begin{aligned}\label{dot2}
\dot{\mathbf{m}}_s  = i\phi_s \mathbf{m} -i\phi\psi\bt - (i(1-\alpha')+\alpha)\psi_{ss}\bt  -\frac{1}{2}(i(1-\alpha')+\alpha)\psi_s (\psi^*\mathbf{m}+\psi\mathbf{m}^*)\,.
\end{aligned}\ee

Clearly, the coefficients of $\bt$, $\mathbf{m}$ and $\mathbf{m}^*$ in equations \eqref{dot1} and \eqref{dot2} should match exactly. The $\mathbf{m}^*$ coefficients already match exactly. Setting the $\mathbf{m}$ coefficients equal, we obtain
\be 
\phi_s = \frac{1-\alpha'}{2}\frac{\pa}{\pa s}|\psi|^2 -\frac{\alpha i}{2}(\psi^*\psi_s - \psi\psi_s^*) \,,
\ee
hence
\be \begin{aligned}\label{phi}
\phi(s,t) & = \frac{1-\alpha'}{2}|\psi|^2 + \alpha (\text{Re}(\psi))(\text{Im}(\psi)) + A(t)\\
& = \frac{1-\alpha'}{2}|\psi|^2  -\frac{i\alpha}{4} (\psi^2 - {\psi^*}^2) + A(t) \,,
\end{aligned}\ee
where $A(t)$ is an arbitrary function of time. Despite the appearance of $i$, this representation is real-valued, since $\psi^2 - {\psi^*}^2$ is purely imaginary. Matching the coefficients of $\bt$, we obtain
\be 
\dot{\psi} +\alpha|\psi|^2\psi = i \phi\psi + (i(1-\alpha')+\alpha)\psi_{ss}\,.
\ee
Using \eqref{phi}, we obtain an evolution equation for the function $\psi$:
\be   \label{eveq}
\dot{\psi}  = iA(t)\psi + (i(1-\alpha')+\alpha)\psi_{ss}  + \left(\frac{i(1-\alpha')}{2}-\alpha\right)|\psi|^2\psi + \frac{\alpha}{4}(\psi^2 - {\psi^*}^2)\psi\,.
\ee
Eq. \eqref{eveq} is a type of complex Ginzburg-Landau equation (GLE) with an additional nonlinearity, $(\psi^2 - {\psi^*}^2)\psi$. If we take $\alpha,\alpha' =0$ (which corresponds to a classical fluid), then \eqref{eveq} reduces to a cubic NLS, and therefore these results are completely consistent with those of Hasimoto for the classical fluid LIA.

For simplicity, we may define a composite parameter $\epsilon=\alpha/(1-\alpha') <<1$ and a function $\hat{A}(T)=A(t)/(1-\alpha')=A(T/(1-\alpha'))/(1-\alpha')$. Let us also introduce the scaled time parameter $T=(1-\alpha')t$. Writing $\psi(s,t)=\sqrt{2}R(s,T)\exp(i\theta(s,T))$, \eqref{eveq} can be written as a real system:
\be \begin{aligned}\label{sys}
R_T & = \epsilon(R_{ss}-R\theta_{s}^2)-2R_s\theta_s -R\theta_{ss}-2\epsilon R^3\,,\\
R\theta_T & = \hat{A}(T)R+R_{ss}-R\theta_s^2 + \epsilon(2R_s\theta_s +R\theta_{ss})  + (1+\epsilon\sin(2\theta))R^3\,.
\end{aligned}\ee

Before moving on, let us remark that the equation \eqref{eveq} is not exactly a complex cubic GLE. While the nonlinearity is of cubic order, the nonlinearity in the complex cubic GLE would take the form $|\psi|^2\psi$. Therefore, the dynamics of \eqref{eveq} may be a bit more complicated than those of the complex cubic GLE, at least in some parameter regimes. Regarding the arbitrary function $A(t)$, note that this function can be picked so that the vortex filament satisfies certain structural conditions. In the case of a solitary wave, picking $A(t)$ to be constant gives $A(t)$ the interpretation of being a spectral parameter. This arbitrary function is sometimes specified in the NLS case so that the solution to an initial value problem has a smooth solution for $s\geq 0$ (that is, so that there is not any mismatch between a boundary condition at $s=0$ and the solution valid on $s>0$). For the classical LIA to NLS map considered in Bianca and Vega (2012), the NLS 
\be 
i\psi_t + \psi_{ss} + \frac{1}{2}(|\psi|^2 - A(t))\psi =0
\ee
(with $\psi$ defined as in \eqref{defn}) was taken to have $A(t)$ satisfying
\be 
A(t) = \pm 2 \frac{\kappa_{ss}(0,t)-\kappa(0,t)\tau(0,t)^2}{\kappa(0,t)}+\kappa(0,t)^2\,,
\ee
hence $A(t)$ was picked to take into account boundary data for the vortex filament configuration. Regarding similarity solutions to the classical LIA, $A(t) = \frac{1}{t}$ was picked in the case where the initial data $\psi(s,0) = ~\text{p.v.}~ \frac{1}{s}$ (Gutierrez and Vega, 2013).

\section{Stokes waves along a quantum vortex filament}
A Stokes wave solution exists for the classical LIA. To recover a Stokes wave along a quantum vortex filament, we assume a solution which strictly depends on the time variable (i.e. $R=R(T)$, $\theta=\theta(T)$). Then \eqref{sys} reduces to
\be \begin{aligned}\label{sysStokes}
R_T & = -2\epsilon R^3\,,\\
R\theta_T & = \hat{A}(T)R + (1+\epsilon\sin(2\theta))R^3\,.
\end{aligned}\ee
Clearly, $R(T) = (1 + 4\epsilon T)^{-1/2}$, so 
\be \label{theta1}
\theta_T  = \hat{A}(T) + \frac{1+\epsilon\sin(2\theta)}{1 + 4\epsilon T}\,.
\ee
If we neglect the $\sin(2\theta)$ term (which is reasonable for small $\epsilon$), we obtain the solution
\be \label{analytic}
\theta(T) =  \int_0^T \hat{A}(\overline{T})d\overline{T} + \frac{\ln(1+4\epsilon T)}{4\epsilon}\,.
\ee
In Fig. 1, we demonstrate that this approximation is a valid approximation to the true dynamics of \eqref{theta1}.

\begin{figure}
\begin{center}
\includegraphics[scale=0.4]{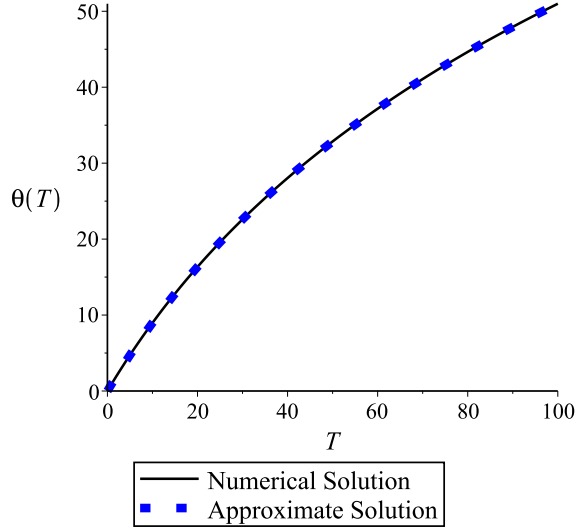}
\end{center}
\caption{(Color Online) Plot of the numerical solutions of \eqref{theta1} along with the approximate analytical solution \eqref{analytic} when temperature 1.0K ($\alpha = 0.006$, $\alpha' = 0.003$), $\hat{A}(T) \equiv 0$, and $\theta(0)=0$. The approximate solution obtained by neglecting the $\epsilon\sin(2\theta)$ term in \eqref{theta1} is in excellent agreement with the numerical solution. \label{Fig1}}
\end{figure}

Note that when $\epsilon =0$, we recover the classical LIA solution of constant modulus, which persists in time. Such a solution would be a vortex ring. However, when $\epsilon \neq 0$, the Stokes waves along a quantum vortex filament necessarily decay in time (with the manner of decay being algebraic of the order $T^{-1/2}$). The dynamics of the equations governing these solutions are dissipative in time. Physically, this shows us exactly how the mutual friction parameters cause the decay of Stokes waves along a quantum filament in the quantum LIA, as expected. In terms of curvature, this solution can be described (using \eqref{defn}) by
\be 
\kappa(s,t) = \sqrt{\frac{2}{1+4\alpha t}}\,.
\ee
The torsion is clearly zero. For the classical LIA, this would be a constant curvature solution. However, with the inclusion of mutual friction, the Stokes wave has decaying curvature and tends towards a zero-curvature solution for large time. As such, constant curvature solutions do not exist for the quantum LIA under non-zero mutual friction, if we assume only time dependent solutions. This means that the classical vortex ring will gradually collapse under mutual friction effects.

What we have shown here is that there is no static vortex ring in the quantum case studied here. In addition to the fact that we do not have a static vortex ring, there are some other classical solutions that do not exist in the quantum case in the presence of mutual friction alone. Indeed, one may show that both a classical helical solution and the planar solution will not exist under the quantum formulation we consider here, for reasons we explain below.

\subsection{No helical filaments under the complex scalar PDE model}
The helical filament configuration can exist when we have a non-zero normal fluid velocity. In this case, the structure can be viewed as a filament with Kelvin waves that propagate along the filament. The normal fluid velocity drives these filaments, as discussed in Van Gorder (2014b). In the case with zero normal fluid velocity, these structures do not exist. This was shown in Van Gorder (2014a) for a type of potential model (where the tangent vector $\mathbf{t}$ was mapped into a complex scalar equation distinct from the curvature-torsion formulation given here), hence it makes sense that the same result would hold here. The helical filament has constant curvature, while its torsion is space and time dependent (in contrast to the vortex ring discussed above, which has constant curvature yet zero torsion). If curvature is constant, we must have $R(s,T)=R_0$, where $R_0$ is some constant. Then the system \eqref{sys} becomes
\be \begin{aligned}\label{syshelix}
  \epsilon \theta_{s}^2  + \theta_{ss}+2\epsilon R_0^2 =0\,,\\
\theta_T - \epsilon \theta_{ss} + \theta_s^2 - \epsilon \sin(2\theta) R_0^2 = A(T) + R_0^2\,.
\end{aligned}\ee
The first equation admits an exact solution, but this solution cannot be made to satisfy the second equation, and hence the system of differential equations is inconsistent. Therefore, there is no solution $\theta(s,T)$ which satisfies \eqref{syshelix}. This in turn implies that there is no helical filament solution to the model we study. The form of a quantum helical filament was given in Van Gorder (2014b), and we see that for such a solution the role of the normal fluid flow is pivotal, as it drives the Kelvin waves along the filament. Therefore, it makes sense that such structures are not found when we set $\mathbf{U}=\mathbf{0}$. Essentially, the normal fluid velocity is needed to balance the dissipative effects due to the mutual friction terms in a precise way on order to allow static helical solutions.

\subsection{No planar filaments under the complex scalar PDE model}
The solutions discussed in this section have thus far involved some assumption on curvature. The analog of the Stokes wave has space independent curvature, while helical filaments or vortex rings have space and time independent curvature (and do not exist here, due to dissipative effects). It is possible to consider a filament with variable curvature yet zero torsion. Such a solution would be planar (in the sense that a three-dimensional curve with zero torsion can be embedded in a plane), and would therefore generalize the classical planar solution due to Hasimoto (1971). Let us assume $\theta(s,T)=\theta(T)$, so that $\tau(s,T) =0$. Then, \eqref{sys} reduces to
\be \begin{aligned}\label{syshelix2}
 R_T - \epsilon R_{ss} + 2\epsilon R^3 =0\,,\\
(\theta_T(T) - A(T) )R - R_{ss} - (1+\epsilon\sin(2\theta(T))) R^3 =0\,.
\end{aligned}\ee
Solving the latter equation, we obtain an expression for $R$ in terms of a Jacobi elliptic function (of type sn) in $s$ with coefficients in $T$. Yet, placing this expression into the first equation, we find that this first equation is never satisfied identically, even if we are free to pick $A(T)$ and $\theta(T)$. As such, there is no zero-torsion (planar) filament solution with non-zero curvature. 

While the above demonstrates mathematically the problem of having a purely planar filament, physically the friction effects would have the effect of inducing torsion on a filament which has zero-torsion at time $t=0$. Therefore, as time increases, what may initially be a planar vortex filament would appear to bend as it rotates about a central axis. That said, there is a way to generalize the planar filament to the quantum case such that when $\alpha,\alpha' \rightarrow 0$, we recover the classical purely planar solution. This solution essentially behaves as we would expect. At order $O(1)$, we have a planar filament which gradually deforms due to torsion effects of order $O(\alpha)$. This construction is discussed in Van Gorder (2014c).

\section{Standing soliton on a quantum vortex filament}
We shall now generalize the Hasimoto 1-soliton solution for the quantum case. Let $\hat{A}(t) = A_0$, $S=\sqrt{\omega -A_0}s$, $\xi = T/(\omega -A_0)$, $R(s,T) = \sqrt{2(\omega -A_0)}q(S,\xi)$, $\theta(s,T) = \omega T + \epsilon\Theta(S,\xi) = \omega(\omega -A_0) \xi + \epsilon \Theta(S,\xi)$. Since $A_0$ just shift the spectral parameter, we set $A_0=0$. Then
\be \label{sol1}
q_{SS} - q + 2q^3 + \epsilon \left\lbrace q\Theta_\xi + 2\sin(2\epsilon\Theta + 2\omega^2\xi)q^3 \right\rbrace = O(\epsilon^2)\,,
\ee
\be \label{sol2}
q_\xi - \epsilon \left\lbrace q_{SS} -2q_{S}\Theta_S -q\Theta_{SS}-4q^3 \right\rbrace = O(\epsilon^2)\,.
\ee
Let us assume that $q(S,\xi) = q_0(S,\xi) + \epsilon q_1(S,\xi) + O(\epsilon^2)$. Equations \eqref{sol1}-\eqref{sol2} give
\be \label{sol3a}
(q_0)_{SS} -q_0 +2q_0^3 =0\,, 
\ee
\be \label{sol3b}
(q_0)_\xi =0\,,
\ee
\be \label{sol3c}
(q_1)_{SS} -q_1 +6q_0^2 q_1 + q_0\Theta_\xi + 2\sin(2\omega^2\xi)q_0^3 =0\,,
\ee
\be \label{sol3d}
(q_1)_\xi = (q_0)_{SS} -2(q_0)_S\Theta_S -q_0\Theta_{SS} -4q_0^3\,.
\ee
From \eqref{sol3a}-\eqref{sol3b}, $q_0(S)=\text{sech}(S)$, so in the $\epsilon \rightarrow 0$ limit, the result corresponds to the Hasimoto soliton. Assume that 
\be 
q_1(S,\xi) = Q_1(S) + Q_2(S)\sin(2\omega^2 \xi)\,,
\ee
\be 
\Theta(S,\xi) = \Theta_1(S) + \Theta_2(S)\cos(2\omega^2\xi)\,.
\ee
Then, \eqref{sol3c}-\eqref{sol3d} imply
\be \label{sol4a}
Q_1''+(6q_0^2 -1)Q_1 =0\,, 
\ee
\be \label{sol4b}
Q_2'' + (6q_0^2 -1)Q_2 - \mu q_0\Theta_2 +2q_0^3 =0\,,
\ee
\be \label{sol4c}
q_0''-4q_0^3 -2q_0'\Theta_1' -q_0\Theta_1'' =0\,,
\ee
\be \label{sol4d}
\mu Q_2 + 2q_0'\Theta_2' + q_0\Theta_2'' =0\,.
\ee
Equations \eqref{sol4a} and \eqref{sol4c} respectively give
\be 
Q_1(S) = \tanh(S)\text{sech}(S)\,,
\ee
\be 
\Theta_1(S) = \frac{3}{4}\left( 1 - \cosh(2S) \right) -2\ln(\cosh(S))\,.
\ee
Unfortunately, equations \eqref{sol4b} and \eqref{sol4d} are coupled and do not admit a clear exact solution in $Q_2$ and $\Theta_2$ if one attempts to solve the equations simultaneously. Note that from \eqref{sol4b} we have
\be \label{theta2a}
\Theta_2(S) = \frac{2}{\mu}\text{sech}^2(S) + \frac{\cosh(S)[Q_2''(S) + (6\text{sech}^2(S) -1)Q_2(S)]}{\mu}\,.
\ee 
Placing this into \eqref{sol4d}, we obtain a forth-order equation for $Q_2$:
\be\begin{aligned} 
 Q_2^{(iv)} + (6-8\tanh^2(S))Q_2'' & - 24\tanh(S)\text{sech}^2(S) Q_2' \\
 & + [4\omega^4 +8\tanh^2(S) -7  + 24\tanh^2(S)\text{sech}^2(S)]Q_2\\
& \qquad = \text{sech}^3(S) \left( 20\tanh^2(S) -4 \right)\,.
\end{aligned}\ee
While a closed form solution to this equation is not forthcoming, we can study the asymptotic behavior of the solution $Q_2(S)$. For large $|S|$, this equation scales as
\be 
Q_2^{(iv)} -2Q_2'' + (4\omega^4 +1)Q_2 =0\,.
\ee
The asymptotics for $Q_2(S)$ are given by
\be \label{asyp1}
Q_2(S) \sim e^{\mp \chi_{+} S}\cos(\chi_{-} S) \quad \text{as} \quad S \rightarrow \pm \infty\,,
\ee
where
\be 
\chi_{\pm} = \frac{1}{\sqrt{2}}\sqrt{ \sqrt{4\omega^4 + 1} \pm 1 }\,.
\ee
Note that the function $Q_2(S)$ has an exponential rate of decay as $S \rightarrow \pm \infty$, and any temporal effects on the amplitude of the soliton are confined to the small-$S$ regime. 

From \eqref{theta2a}, and using \eqref{asyp1}, we have the asymptotics for $\Theta_2(S)$:
\be 
\Theta_2(S) \sim e^{\mp \chi_{+} S}\cosh(S)\cos(\chi_{-} S) \sim e^{\pm(1 - \chi_{+}) S}\cos(\chi_{-} S)\,,
\ee
as $S \rightarrow \pm \infty$. Since $\chi_+ \geq 1$, the function $\Theta_2(S)$ does not grow as $S \rightarrow \pm \infty$. This means that the terms in $\Theta(S,\xi)$ associated with $\Theta_1(S)$ dominate. 

Putting these results back into natural coordinates, we find that the curvature is given by
\be 
\kappa(s,t) = 2\sqrt{\omega}\left\lbrace \text{sech}(\sqrt{\omega}s)\left\lbrace 1 + \frac{\alpha}{1-\alpha'}\tanh(\sqrt{\omega}s) \right\rbrace  + \frac{\alpha}{1-\alpha'}Q_2(\sqrt{\omega}s)\sin(2\omega(1-\alpha')t) \right\rbrace\,,
\ee
while the torsion is given by
\be 
\tau(s,t) = -\frac{\alpha \sqrt{\omega}}{1-\alpha'}\left\lbrace \frac{3}{2}\sinh(2\sqrt{\omega}s) + 2\tanh(\sqrt{\omega}s) - \Theta_2'(\sqrt{\omega}s)\cos(2\omega(1-\alpha')t)  \right\rbrace\,.
\ee
For large $|s|$, the magnitude of the torsion becomes arbitrarily large. However, in the same limit, the curvature tends rapidly to zero. So, while the twisting of the filament becomes arbitrarily large, this segment of the filament will look like a line. Near the origin, curvature is maximized while the magnitude of the torsion is minimal. In this region, we obtain a stationary topological defect. 

In Fig. 2, we plot the curvature $\kappa(s,t)$. The time-dependent perturbations are not evident at the scale (order $O(1)$) of the soliton, so in a separate graph we plot $\kappa(s,0)-\kappa(s,\overline{t})$ where $\overline{t} = \frac{\pi}{4(1-\alpha')}$. This gives us a good understanding of the order $O(\alpha)$ perturbations. The form of these perturbations changes with time in a periodic manner, and these perturbations essentially make the soliton envelope wavy.

\begin{figure}
\begin{center}
\includegraphics[scale=0.3]{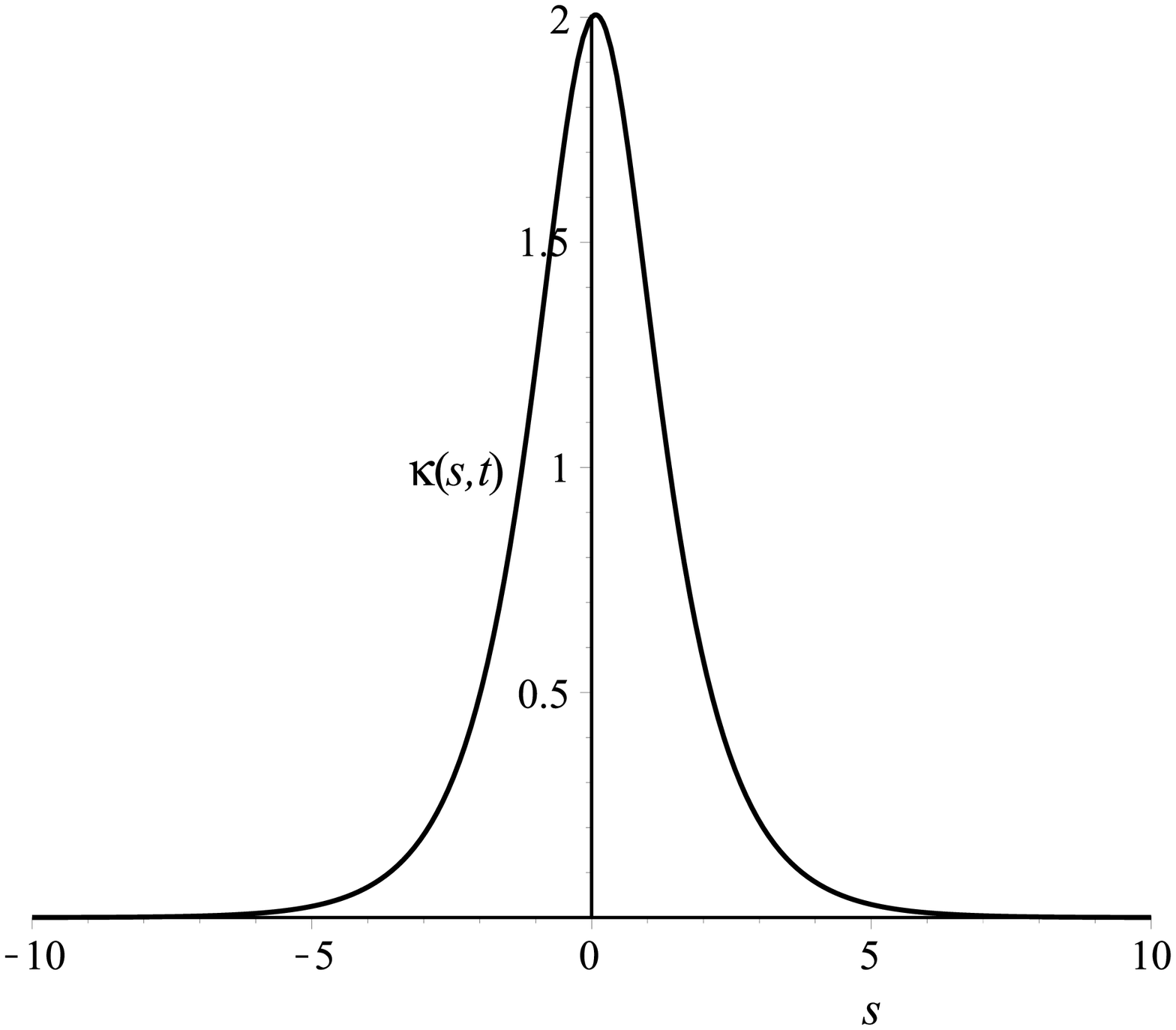}\includegraphics[scale=0.3]{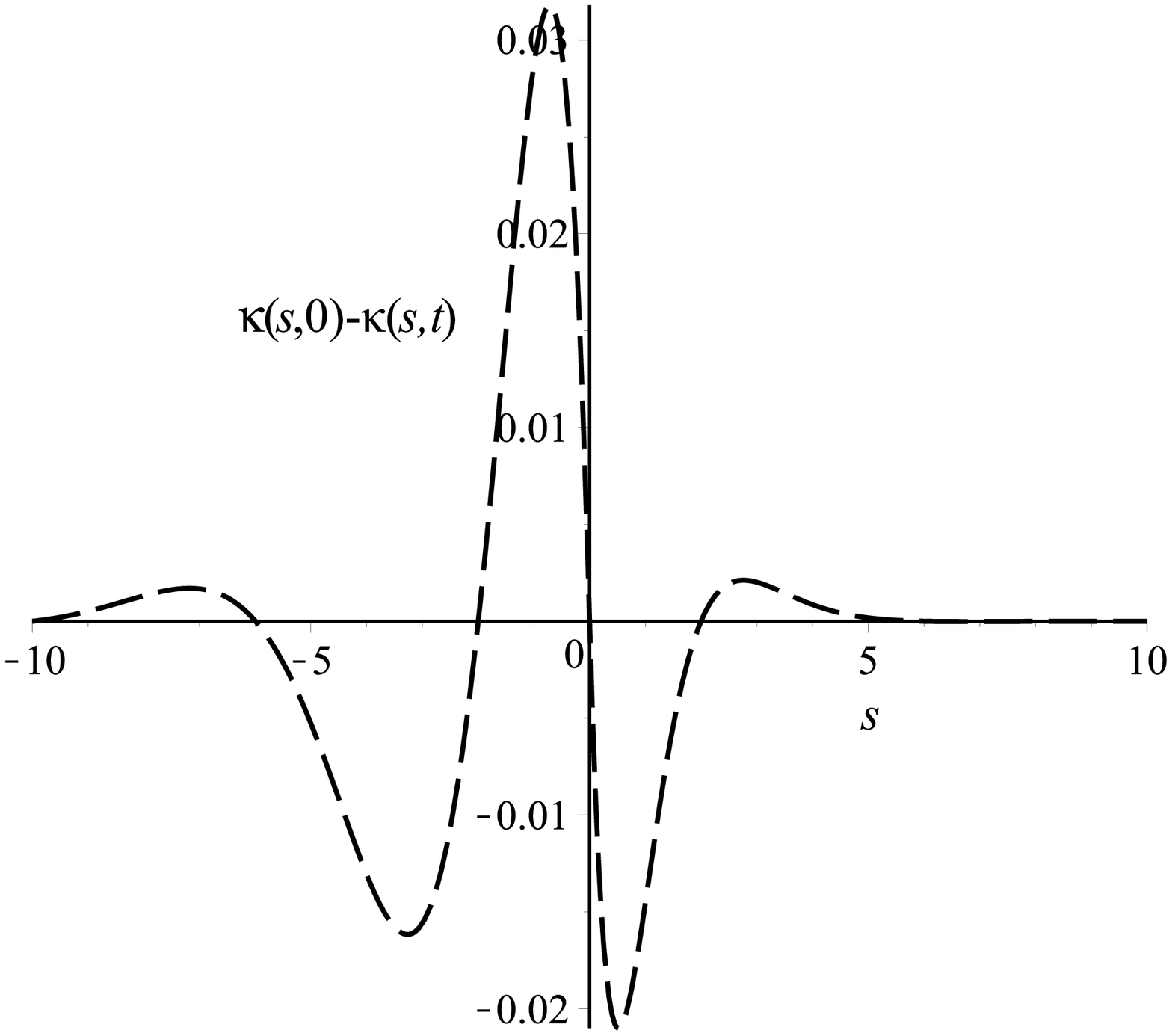}
\end{center}
$$
(a) \qquad\qquad\qquad\qquad\qquad\qquad\quad (b)
$$
\caption{Plot of (a) the soliton solution $\kappa(s,t)$ at $t=\frac{\pi}{4(1-\alpha')}$ and (b) the order $O(\alpha)$ perturbations (once the leading $2\text{sech}(s)$ term is removed) at $t=\frac{\pi}{4(1-\alpha')}$. The plots correspond to the temperature 1.5K regime ($\alpha = 0.073$, $\alpha' = 0.018$), while the spectral parameter is set to $\omega =1$. Note that since there is no apparent exact closed form solution for $Q_2$, we solve for it numerically when obtaining these plots. Similar plots are found for the 1.0K temperature regime, with the only difference being that the $O(\alpha)$ perturbations are necessarily smaller. \label{Fig2}}
\end{figure}

This soliton solution is interesting for a few reasons. First of all, the soliton solution to the classical LIA obtained by Hasimoto (1972) motivated a number of studies on vortex filament dynamics. For instance, Hopfinger and Browand (1982) studied turbulent flow in a rotating container, and in particular observed waves along vortex filaments. They determined that the theoretical solutions predicted by Hasimoto (1972) do actually occur along vortex filaments in practical experiments, which in turn motivated even further work. It therefore makes sense that solitons, such as the one obtained here from the quantum LIA, should occur along quantum filaments, and that such solutions should be very similar in form to their classical counterparts. The stationary Hasimoto 1-soliton is torsionless (although for non-stationary waves, this is not in general the case) while the stationary solution (in the small-time regime) does have non-zero torsion, and this is the primary qualitative difference between the solution given here and the standing wave of Hasimoto.

It is interesting that solitons emerge from the transformed model \eqref{eveq}, since the model is dissipative. However, the solitons are valid for small timescales, and dissipative effects likely dominate any nonlinear effects for large time. We should note that solitons have indeed been found in variations of complex GLEs in the past; for a review, see Aranson and Kramer (2002). The dissipative effects (the influence of mutual friction terms) are confined in their influence on the phase term (the torsion) rather than the amplitude (the curvature). This has the result of inducing torsion effects in space due to the presence of mutual friction terms whereas curvature remains independent of this mutual friction. Physically, we conclude that standing solitons along a quantum vortex filament do arise from the quantum LIA, and the primary difference in these solitons from their classical counterparts (which have been described theoretically (Hasimoto, 1972) and experimentally (Hopfinger and Browand, 1982)) is that the mutual friction terms induce a stronger spatial dependence on the torsion of the curve defining the vortex filament (as mentioned above, even from the standing wave, there will be torsion effects, which is distinct from the classical case). Hence, not only do solitons exist in the quantum form of the LIA for small timescales, but these solitons are exactly the generalization of the classical Hasimoto-type solitons which account for superfluid dissipation effects. For larger timescales, one should expect a gradual dissipation of the amplitude of the solitons (the curvature of the filament), meaning that for large time the soliton along the filament should gradually decay, with the quantum filament taking the appearance of a line filament in this limit.

To summarize, the soliton found in this section has the following properties:\\
(i) The overall soliton envelope scales as $|\psi(s,t)| \sim \text{sech}(s) + O(\alpha)$ and is similar in form to that of the classical case. When $\alpha \rightarrow 0$, we recover the classical soliton of Hasimoto.\\
(ii) The first order correction to the curvature involves a periodic bounded function of time, as well as a bounded function of arc-length. Hence, the first order corrections will exhibit small oscillations around the time-averaged envelope
$$
|\psi(s,t)|_{\text{avg}} = 2\sqrt{\omega} \text{sech}(\sqrt{\omega}s)\left\lbrace 1 + \frac{\alpha}{1-\alpha'}\tanh(\sqrt{\omega}s) \right\rbrace 
$$
as time increases. \\
(iii) The torsion of the quantum filament also depends on a periodic function of time. However, for large $|s|$, the time-dependent term decays, while the strictly space dependent term dominates.\\
(iv) While torsion increases away from the origin, curvature rapidly decays for large $|s|$ and therefore the filament takes on the appearance of a line sufficiently far from the origin. \\
(v) Any dissipative effects on the amplitude of the soliton are of order $O(\alpha^2)$ or greater. This means that dissipation of this soliton is expected to be slow.

\section{Self-similar waves}
Previous studies have considered self-similar solutions for the LIA in both the classical (Gutierrez, Rivas and Vega, 2003) and quantum (Lipniacki, 2003a,b; Van Gorder, 2013a) cases. Results for the quntum LIA were previously reported for the curvature and torsion directly and also for the Cartesian representation (the latter in the small-amplitude regime).

Let us write $\psi(s,T) = \frac{1}{\sqrt{T}}\Psi(\sigma)$ where $\sigma = s/\sqrt{T}$ is the similarity variable. We then obtain
\be 
(i+\epsilon)\Psi'' + \frac{\sigma}{2}\Psi' + \frac{1}{2}\Psi + \left( \frac{i}{2}-\epsilon \right)|\Psi |^2\Psi + \frac{\epsilon}{4}\left( \Psi^2 - {\Psi^*}^2 \right)\Psi =0\,.
\ee
Upon writing $\Psi(\sigma)=\sqrt{2}\rho(\sigma)\exp(i\phi(\sigma))$, \eqref{eveq} is reduced into a system of real ordinary differential equations:
\be \label{s5}
\epsilon (\rho'' -\rho{\phi'}^2) -2\rho'\phi' -\rho\phi'' +\frac{\sigma}{2}\rho' + \frac{1}{2}\rho -2\epsilon\rho^3 =0\,,
\ee
\be \label{s6}
\rho'' -\rho{\phi'}^2 +2\epsilon\rho'\phi' +\epsilon\rho\phi'' +\frac{\sigma}{2}\rho\phi' + (1+\epsilon\sin(2\phi))\rho^3 = 0\,.
\ee

A solution with small curvature will approximately satisfy
\be \label{simple}
(i+\epsilon)\Psi'' + \frac{\sigma}{2}\Psi' + \frac{1}{2}\Psi  =0\,.
\ee
Consider the conditions $\Phi(0)=\delta <<1$ and $\Phi'(0)=0$. The solution of \eqref{simple} is then
\be 
\Psi(\sigma) = \delta \exp\left( \frac{i-\epsilon}{4(1+\epsilon^2)}\sigma^2 \right)\,.
\ee
In terms of the natural variables, we have
\be 
\psi(s,T) = \frac{\delta}{\sqrt{T}}\exp\left( \frac{i-\epsilon}{4(1+\epsilon^2)}\frac{s^2}{T} \right)\,.
\ee
Curvature and torsion as given by \eqref{defn} are found to be
\be 
\kappa(s,T) = \frac{\delta}{\sqrt{T}}\exp\left( -\frac{\epsilon}{4(1+\epsilon^2)}\frac{s^2}{T} \right)\,,
\ee
\be 
\tau(s,T) = \frac{1}{2(1+\epsilon^2)}\frac{s}{T}\,.
\ee
At any fixed value of time, the curvature is clearly maximal at $s=0$. On the other hand, if we focus on some point on the filament corresponding to $s=s_*$, then the curvature at this point is maximal at time
\be 
T_* = \frac{\epsilon {s_*}^2}{2(1+\epsilon^2)}\,.
\ee
So, different points on the vortex filament experience their maximal curvature at different times, so there is not a globally uniform decay or growth of curvature as time increases. On the other hand, the torsion uniformly decreases for any fixed $s$ as time increases. In the limit $T\rightarrow \infty$, torsion tends to zero for any fixed point on the filament, meaning that in the large-time limit the solution appears nearly planar. Meanwhile, curvature tends to zero in the limit $T\rightarrow \infty$, as well. What these facts imply is that the small-curvature solution shown here will degenerate into a line filament for large enough values of time.

\begin{figure}
\begin{center}
\includegraphics[scale=0.4]{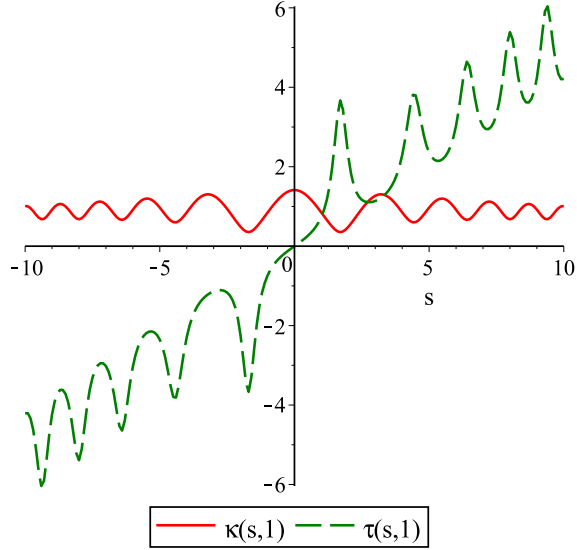}
\end{center}
\caption{(Color Online) Similarity solution plots which include the curvature $\kappa(s,T)=\sqrt{\frac{2}{T}}\rho(\sigma)$ and torsion $\tau(s,T)=\frac{1}{\sqrt{T}}\phi'(\sigma)$ over $s$ when the temperature is 1.0K ($\alpha = 0.006$, $\alpha' = 0.003$), for fixed time $T=1$. Initial conditions are $\rho(0)=1$, $\rho'(0)=0$, $\phi(0)=0$, $\phi'(0)=0$. As time increases, the curvature shown here would gradually flatten and decrease in magnitude.  \label{Fig3}}
\end{figure}

\begin{figure}
\begin{center}
\includegraphics[scale=0.4]{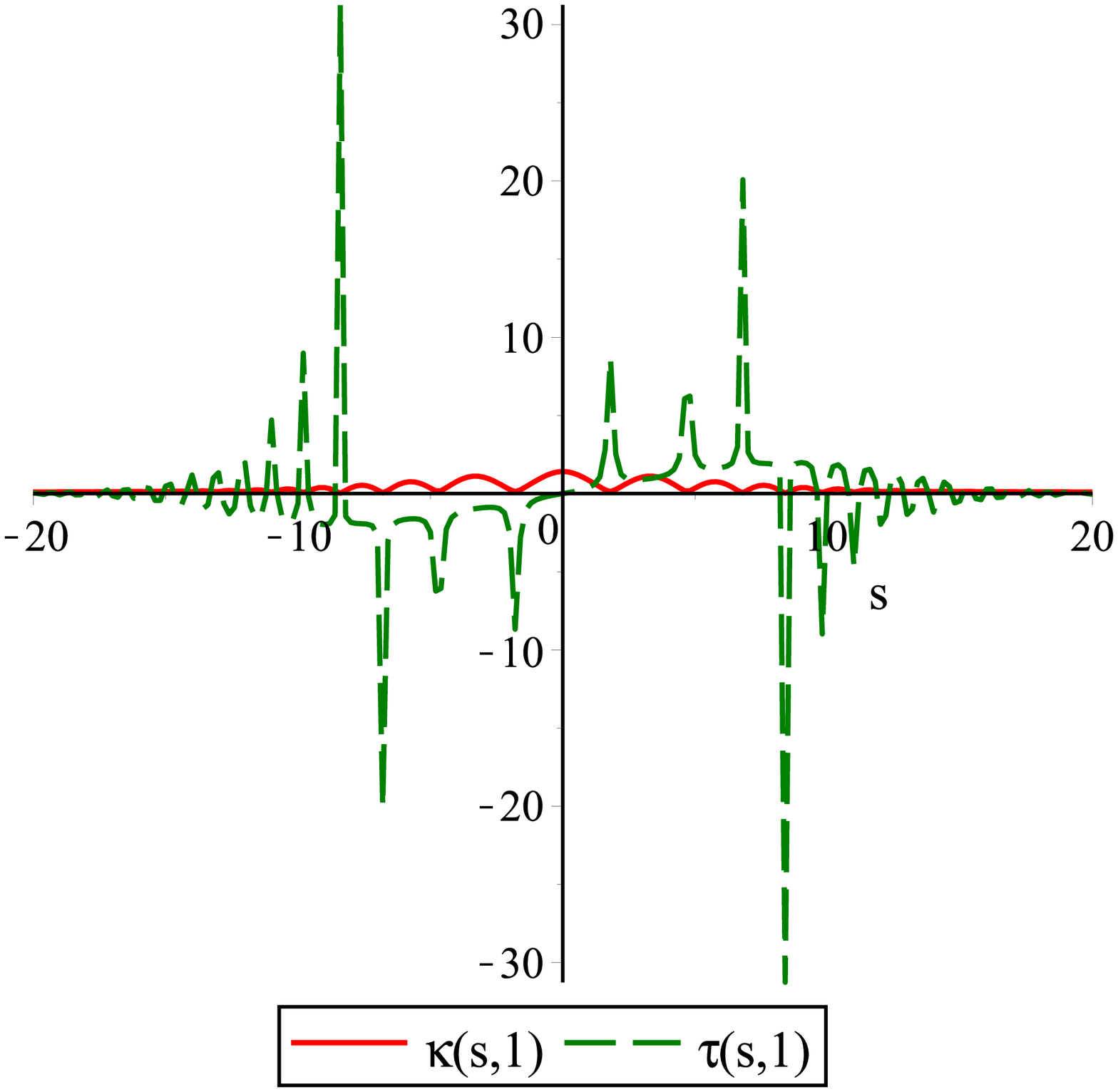}
\end{center}
\caption{(Color Online) Similarity solution plots which include the curvature $\kappa(s,T)=\sqrt{\frac{2}{T}}\rho(\sigma)$ and torsion $\tau(s,T)=\frac{1}{\sqrt{T}}\phi'(\sigma)$ over $s$ when the temperatue is 1.5K ($\alpha = 0.073$, $\alpha' = 0.018$), for fixed time $T=1$. Initial conditions are $\rho(0)=1$, $\rho'(0)=0$, $\phi(0)=0$, $\phi'(0)=0$. Again, as time increases, the curvature shown here would gradually flatten and decrease in magnitude. Compared with Fig. 3, the ``warmer" solution here exhibits less regularity (in the torsion function). However, solutions here decay eventually for large $s$, while the torsion increases (in average magnitude) as $|s|$ increases in the cooler superfluid case of Fig. 3. Of course, for large enough time, the curvature and torsion will both decrease at a rate $1/\sqrt{T}$.  \label{Fig4}}
\end{figure}

For more general types of self-similar solutions, we may numerically solve the system \eqref{s5}-\eqref{s6}. From these solutions, we recover $\kappa(s,T)=\sqrt{\frac{2}{T}}\rho(\sigma)$ and $\tau(s,T)=\frac{1}{\sqrt{T}}\phi'(\sigma)$. In Figs. 3-4 we plot solutions corresponding to temperatures 1.0K and 1.5K, respectively. For sake of generating plots, we take $T=1$ and plot the solutions over $s$. These solutions gradually broaden and decrease in magnitude as time increases. For large enough time, curvature eventually tends to zero (as was discussed above), and the solution tends to a line filament away from a neighborhood surrounding the origin $s=0$. For the cooler superfluid (temperature equal to 1.0K), the torsion increases in magnitude (in an average sense). When the superfluid warms (temperature near 1.5K), we see less regularity in the torsion, although for large enough $s$ the torsion tends toward zero. Therefore, the solution appears planar sufficiently far away from the origin. This is in agreement with some of the solutions found previously (Lipniacki, 2003a,b; Van Gorder, 2013a), where interesting behaviors (sharp turns or kinks) are found near the origin yet the filament looks like a line filament far from the origin. For large time, the $1/\sqrt{T}$ factors dominate, and the solutions do decay eventually, so any sharp turns or excitations are gradually smoothed.

\section{Traveling waves}
If we assume that a class of waves along the vortex filament propagate with wave speed $c$, we should look for solutions of the form $R(s,T) = R(z)$ and $\theta(s,T)=\theta(z)$, where $z=s-cT$. Take $\hat{A}(T)=A_0$. The system \eqref{sys} then gives 
\be \label{wave1a}
R'' +A_0 R + R^3 + cR\theta' - R{\theta'}^2 + \epsilon\left\lbrace \sin(2\theta)R^3 +2R'\theta' +R\theta'' \right\rbrace =0\,,
\ee
\be \label{wave1b}
cR' -2R'\theta' -R\theta'' + \epsilon \left\lbrace R'' - 2R^3 -R{\theta'}^2 \right\rbrace =0\,.
\ee

Before proceeding, let us take a moment to consider the classical case. Taking $\epsilon =0$, we recover
\be \label{wave2a}
R'' +A_0 R + R^3 + cR\theta' - R{\theta'}^2  =0\,,
\ee
\be \label{wave2b}
cR' -2R'\theta' -R\theta'' =0\,.
\ee
Equation \eqref{wave2b} is separable, with
\be 
\frac{R'}{R} = \frac{\theta''}{c-2\theta'}\,.
\ee
Integrating once,
\be 
\ln(R) + \frac{1}{2}\ln(c-2\theta') = k_0\,, 
\ee
where $k_0$ is an integration constant. Then,
\be 
R\sqrt{c-2\theta'} = e^{k_0} = k\,,
\ee
or 
\be \label{thetawave}
\theta' = \frac{c}{2} - \frac{k}{2}R^{-2}\,.
\ee
This gives the torsion of the filament under the traveling wave assumption.
Using this representation in \eqref{wave2a}, we obtain
\be \label{preintegral}
R'' + \left( A_0 + \frac{c^2}{4} \right)R + R^3 - \frac{k^2}{4R^3} =0\,.
\ee
This equation admits a first integral, and we find that
\be \label{integral}
{R'}^2 + \left( A_0 + \frac{c^2}{4} \right)R^2 + \frac{1}{2}R^4 + \frac{k^2}{4R^2} = I\,,
\ee
where $I$ is another constant of integration. For appropriate choice of $I$, this first order differential equation admits closed orbits in the $(R,R')$-plane, and hence there exist bounded and periodic solutions of \eqref{integral}. In fact, there are such solutions which satisfy $R(z)>0$ for all $z\in \mathbb{R}$, hence there are positive bounded periodic solutions. This positivity is necessary, in light of the representation \eqref{thetawave}. In Figure 5, we plot both the curvature and the torsion for these solutions corresponding to the classical limit. For comparison, we also give plots for the temperature 1.0K regime.

\begin{figure}
\begin{center}
\includegraphics[scale=0.3]{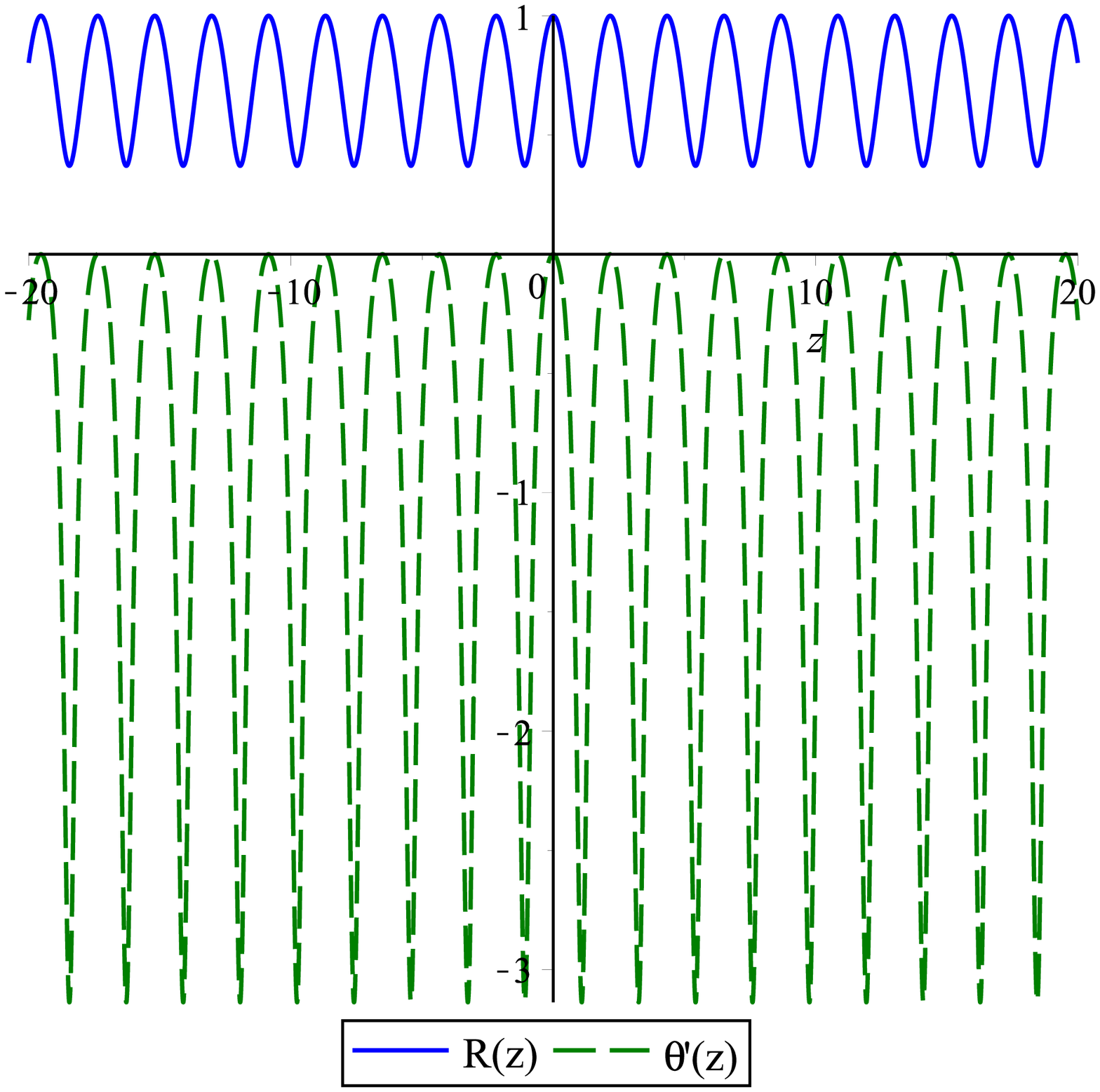}\includegraphics[scale=0.3]{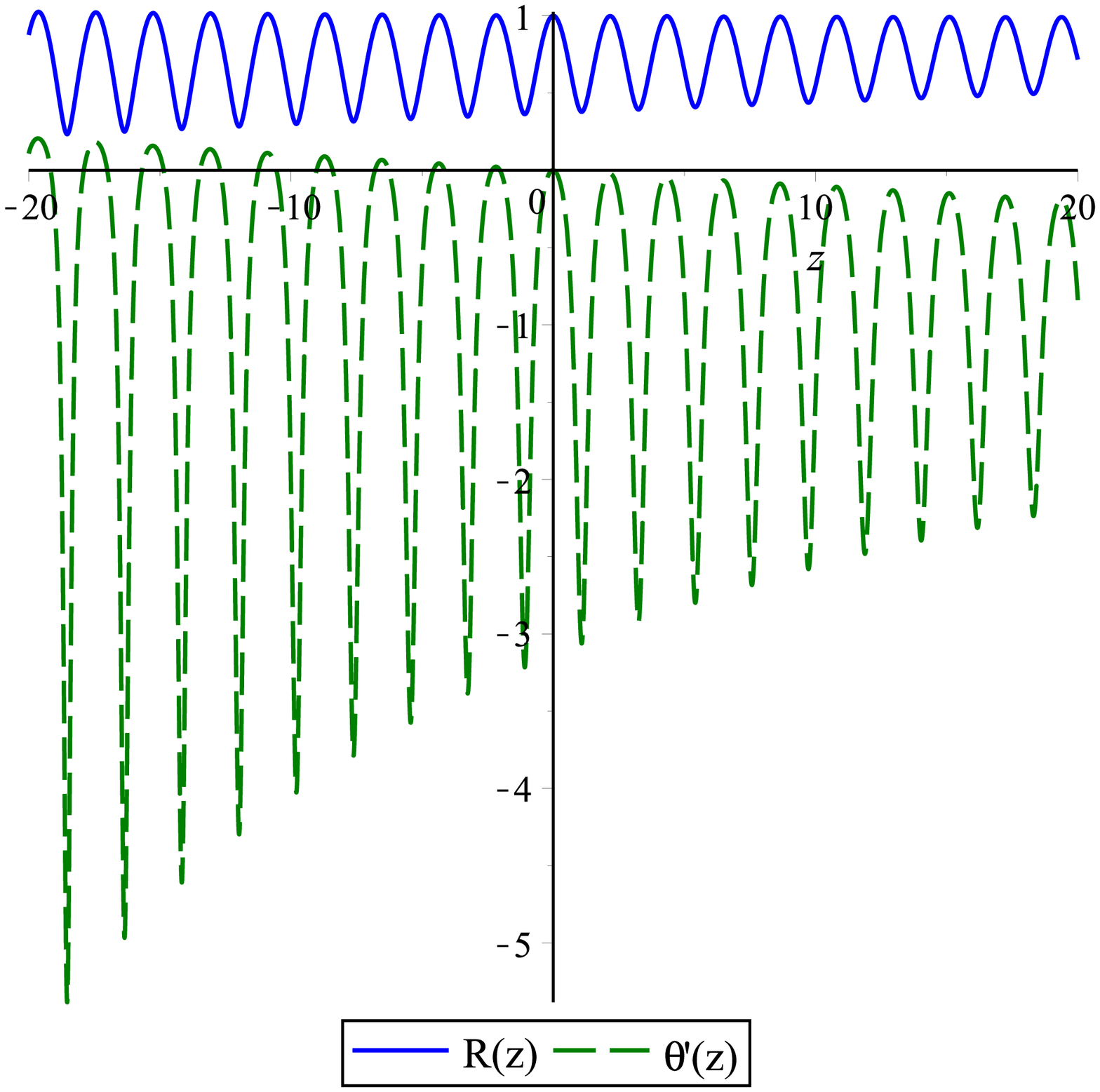}
\end{center}
$$
(a) \qquad\qquad\qquad\qquad\qquad\qquad\quad (b)
$$
\caption{(Color Online) Traveling wave solutions for (a) the classical ($\alpha = \alpha' =0$) and (b) the quantum (temperature 1.0K, $\alpha =0.006$, $\alpha' = 0.003$) cases. The classical solutions are periodic, whereas the quantum solutions lose this periodicity. As predicted from the analysis of the first integral \eqref{integral}, the curvature term $R(z)$ remains bounded and positive. \label{Fig5}}
\end{figure}

With properties of the classical solution in mind, let us return to the quantum case. Equations \eqref{wave1a}-\eqref{wave1b} are essentially perturbations of the equations \eqref{wave2a}-\eqref{wave2b} which account for the influence of mutual friction. It then makes sense to search for solutions of the form 
\be 
R(z) = R_0(z) + \epsilon R_1(z) \qquad \text{and} \qquad \theta(z) = \theta_0(z) + \epsilon \theta_1(z)\,,
\ee
where $R(z)$ is the periodic function described by a solution to \eqref{preintegral} and $\theta_0(z)$ is the unique solution to \eqref{thetawave} subject to $\theta(0)=0$. The perturbation terms are governed by the equations
\be \label{wave3a}
R_1'' + (3R_0^2 +A_0 +c\theta_0'-{\theta_0'}^2)R_1 +R_0(c-2\theta_0')\theta_1' = -\sin(2\theta_0)R_0^3 -R_0\theta_0'' -2R_0'\theta_0'\,,
\ee
\be \label{wave3b}
(c-2\theta_0')R_1' - \theta_0'' R_1 - 2R_0'\theta_1' - R_0\theta_1'' = - R_0'' + R_0{\theta_0'}^2 + 2R_0^3\,.
\ee
While these equations are linear, the form of the coefficients can cause numerical approaches to breakdown, leading to an artificial blow-up of solutions. (Similar issues are found if one attempts to solve the fully nonlinear problem \eqref{wave1a}-\eqref{wave1b} directly.) For large values of $|s|$, the perturbation solutions will break down, which suggests that a more involved perturbation approach might be warranted in order to describe the effect of the $\epsilon \neq 0$ perturbations on the unperturbed equations \eqref{wave2a}-\eqref{wave2b}.

\section{Waves driven by the normal fluid: a potential function formulation}
In the previous sections of this paper, we assumed that normal fluid effects were small in order to derive a quantum analogue of the classical Hasimoto transformation, which puts the quantum LIA into direct correspondence with a scalar nonlinear partial differential equation. We were then able to study a number of quantum vortex filament configurations by obtaining solutions to this partial differential equation, including decaying Stokes waves, a standing soliton, self-similar solutions, and traveling waves. However, we also highlighted the fact that certain other waves (such as Kelvin waves along a quantum filament) are driven by the normal flow, and therefore could not be considered under the quantum Hasimoto transformation (since that transform necessarily exists only when normal flow effects are small enough to be negligible). In the present paper, we shall discuss a different transformation, valid in the case of traveling waves, which preserves the normal fluid velocity (and hence can account for cases where this velocity is large). This transform is quite different from the quantum Hasimoto transformation, and has more in common with the transformation for the tangent vector derived in Van Gorder (2014a).

Let us assume that a wave along the quantum vortex filament modeled by \eqref{fil} depends on a traveling wave variable $z=s-ct$, where $c$ is the wave speed. Note that if $c=0$, we then have a standing wave. Then, equation \eqref{unitLIA} for the tangent vector to the quantum filament takes the form
\be \label{nf1}
-c\bt ' = \frac{d}{dz}\left\lbrace (1-\alpha')\bt \times \bt ' + \alpha \bt ' + \alpha \bt \times \mathbf{U} - \alpha' (\bt \cdot \mathbf{U})\bt \right\rbrace \,.
\ee
where prime denotes differentiation with respect to $z$. Equation \eqref{nf1} is exactly integrable, so integrating \eqref{nf1} once gives the first order vector ordinary differential equation
\be \label{nf2} 
(1-\alpha')\bt \times \bt ' + \alpha \bt ' + \alpha \bt \times \mathbf{U} - \alpha' (\bt \cdot \mathbf{U})\bt + c \bt = \mathbf{I}\,,
\ee
where $\mathbf{I}$ is a constant vector. Let us pick the alignment of $\bt$ so that $\mathbf{U}$ is directed along the first component of $\bt$. Let us write $\bt(z)=[t_1(z),t_2(z),t_3(z)]^T$ and then $\mathbf{U}=[U,0,0]^T$. Calculating the needed quantities in \eqref{nf2}, we have
\be \label{nf3a}
(1-\alpha')(t_2t_3'-t_2't_3) + \alpha t_1' - \alpha' U t_1^2 + ct_1 = I_1\,,
\ee
\be \label{nf3b} 
(1-\alpha')(t_1't_3-t_1t_3') + \alpha t_2' + \alpha U t_3 - \alpha' U t_1t_2 + ct_2 = I_2\,,
\ee
\be \label{nf3c} 
(1-\alpha')(t_1t_2'-t_1't_2) + \alpha t_3' - \alpha U t_2 - \alpha' U t_1t_3 + ct_3 = I_3\,.
\ee
Let us take $i$\eqref{nf3b} - \eqref{nf3c}:
\be \label{nf4}
(1-\alpha')(-t_1(t_2 + it_3)' + t_1'(t_2+it_3)) + \alpha (it_2 -t_3)' + \alpha U (t_2 + it_3) - \alpha' Ut_1 (it_2 -t_3) + c(it_2 -t_3) = iI_2 + I_3\,.
\ee
Let us define the complex quantity $\beta(z) = t_2(z) + it_3(z)$. Then, \eqref{nf4} becomes
\be \label{nf4b}
(1-\alpha')(-t_1\beta' + t_1'\beta) + \alpha i \beta' + \alpha U \beta - \alpha' U i t_1 \beta + c i \beta = I_0\,,
\ee
where $I_0 = iI_2 - I_3$ is a complex constant. Note that since we take $\bt$ to be a unit vector, we should have $t_1(z) = \sqrt{1-t_2(z)^2-t_3(z)^2}=\sqrt{1-|\beta|^2}$. This choice satisfies the remaining condition \eqref{nf3a} (under an appropriate choice of $I_1$; we shall say more on this later), and noting 
\be 
-t_1\beta' + t_1'\beta = -t_1^2 \left( \frac{\beta}{t_1}  \right)' = - \left( 1- |\beta|^2 \right)\left( \frac{\beta}{\sqrt{1-|\beta|^2}} \right)' \,,
\ee
equation \eqref{nf4b} becomes
\be \label{nf5}
(1-\alpha')\left(1-|\beta|^2\right)\left( \frac{\beta}{\sqrt{1-|\beta|^2}} \right)' - \alpha i \beta' - U\left\lbrace \alpha - \alpha' i \sqrt{1-|\beta|^2} \right\rbrace \beta - c i \beta = I_0\,.
\ee
Assuming we can solve \eqref{nf5} for the complex scalar function $\beta$, we can recover the tangent vector by
\be \label{nf5b}
\bt(z) = \left[ \sqrt{1-|\beta(z)|^2}, \text{Re}~\beta(z), \text{Im}~\beta(z) \right]^T \,.
\ee
While there may exist solutions for which $I_0 \neq 0$ is useful, in what follows we set $I_0 =0$ and $I_1$ is set so that \eqref{nf3a} is satisfied.

This formulation involves solving equation \eqref{nf5} for the scalar potential function $\beta(z)$, and then placing this function back into \eqref{nf5b} to recover the tangent vector. These wave solutions essentially depend on two free components, $t_2(z)$ and $t_3(z)$, which in turn fix $t_1(z)$. As we shall see, this formulation is most useful in cases where components $t_2(z)$ and $t_3(z)$ are symmetric in some way, so that the vortex filament is essentially aligned along the first component $t_1(z)$. This corresponds to the direction of the normal fluid velocity $\mathbf{U}$. We can view solutions \eqref{nf5b} as driven or generated by this normal fluid flow, in contrast to the solutions in previous sections of this paper, which were independent of the normal fluid velocity (since this velocity was assumed to be negligibly small). Note that by integrating \eqref{nf5b} in arclength, one may recover the position vector giving the vortex filament curve at any location $s$ and time $t$. In what follows, we give specific examples of solutions $\beta(z)$ to \eqref{nf5}, which we then use to construct the tangent vector as in \eqref{nf5b}.

\subsection{Normal fluid driven helical filament}
Let us take $\beta(z) = A\exp(i\nu z)$, where $A$ and $\nu$ are parameters. Then, \eqref{nf5} reduces to the algebraic equation
\be 
\left\lbrace ((1-\alpha')\nu + \alpha' U)\sqrt{1-A^2}-c\right\rbrace i + \alpha (\nu -U) =0\,.
\ee
Separating real and imaginary parts, we find $\nu = U$ and $c = \sqrt{1-A^2}U$. Therefore,
\be \label{helixagain}
\beta(z) = A \exp\left( i Uz \right) = A \exp\left( i U \left\lbrace  s - \sqrt{1-A^2}U t\right\rbrace \right)\,.
\ee
We then recover the tangent vector
\be 
\bt(z) = \left[ \sqrt{1-A^2}, A\cos\left( Us - \sqrt{1-A^2}U^2 t \right) , -A\sin\left( Us - \sqrt{1-A^2}U^2 t \right) \right]^T \,.
\ee
We must require that $0<A<1$, but otherwise all other parameters are free. When $U\rightarrow 0$, this solution degenerates into a line filament. Therefore, this particular vortex filament only exists when we have a non-zero normal fluid velocity.

\subsection{Normal fluid generated standing wave}
Let us next consider a solution of the form $\beta(z) = \exp(i\nu z)f(z)$, where $f(z)$ is a real-valued function to be determined. After a few manipulations, \eqref{nf5} reduces to
\be 
(1-\alpha')\frac{f'}{\sqrt{1-f^2}} + \alpha(\nu -U)f + \left\lbrace ((1-\alpha')\nu +\alpha' U)f\sqrt{1-f^2} -\alpha f' -cf \right\rbrace i =0\,.
\ee
Now, $f(z)$ is assumed to be real valued, meaning that the equations
\be \label{nf6a}
\frac{f'}{\sqrt{1-f^2}} + \frac{\alpha(\nu -U)}{1-\alpha'}f =0\,,
\ee
\be \label{nf6b}
((1-\alpha')\nu +\alpha' U)f\sqrt{1-f^2} -\alpha f' -cf =0\,.
\ee
must be satisfied simultaneously. Note that \eqref{nf6a} has the exact solution
\be \label{nf6c}
f(z) = \text{sech}\left( \frac{\alpha(\nu -U)}{1-\alpha'} z \right)\,.
\ee
Placing the solution \eqref{nf6c} into \eqref{nf6b}, we find
\be 
\nu = -\frac{\alpha' - \alpha'^2 -\alpha^2}{(1-\alpha')^2 + \alpha^2} U \quad \text{and} \quad c =0
\ee
must hold. This implies that the only exact solution of this type is stationary in time, meaning we have a standing wave. Thus, $z=s$ in this case. This solution therefore takes the form
\be 
\beta(s) = \exp\left( - i \frac{\alpha' - \alpha'^2 -\alpha^2}{(1-\alpha')^2 + \alpha^2} U s\right)\text{sech}\left( \frac{\alpha U}{(1-\alpha')^2 + \alpha^2} s \right)\,.
\ee
We then obtain the tangent vector
\be \begin{aligned}
\bt(s) & = \left[ \tanh\left( \frac{\alpha U}{(1-\alpha')^2 + \alpha^2} |s| \right) , \cos\left(\frac{\alpha' - \alpha'^2 -\alpha^2}{(1-\alpha')^2 + \alpha^2} U s\right)\text{sech}\left( \frac{\alpha U}{(1-\alpha')^2 + \alpha^2} s \right), \right. \\
& \qquad\qquad\qquad \left. -\sin\left(\frac{\alpha' - \alpha'^2 -\alpha^2}{(1-\alpha')^2 + \alpha^2} U s\right)\text{sech}\left( \frac{\alpha U}{(1-\alpha')^2 + \alpha^2} s \right) \right]^T \,.
\end{aligned}\ee
While this structure corresponds to a standing wave, and hence it is not really driven by the normal flow, the existence of this structure depends completely on $U\neq 0$. This structure takes the form of a topological soliton which is stationary in time (although other structures are certainly possible). Despite the fact that the solution is stationary, it has a structure which depends on the normal fluid velocity in a rather fundamental way: if $|\mathbf{U}| \rightarrow 0$, then the solution collapses to a point. Similarly, we need $\alpha,\alpha' >0$. Such a solution is therefore not possible under the classical LIA with $\alpha,\alpha' \rightarrow 0$. This is in contrast to other types of solutions (solitons, helices, and so on) which appear under both the classical and quantum forms of the LIA. This type of solution is therefore one example of a distinct kind of filament which can be observed only under the quantum LIA.

\subsection{Consistency equation for the scalar equation \eqref{nf5}}
Note that equation \eqref{nf5} allows one to solve for a complex scalar function $\beta(z)$, which can then be used to immediately find the components $t_2(z)$ and $t_3(z)$, and since we assume $\mathbf{t}(z)$ is a unitvector, one also finds $t_1(z)=\sqrt{1-t_2(z)^2 - t_3(z)^2}$. However, \eqref{nf5} uses information from equations \eqref{nf3b} and \eqref{nf3c}, while we did not use \eqref{nf3a} in the derivation of \eqref{nf5}. Therefore, one must be sure that an obtained solution $\beta(z)$ to \eqref{nf5} is consistent with \eqref{nf3a} in order for it to correspond to a solution of the vector traveling wave form of the quantum LIA given by \eqref{nf2}. In this subsection, we derive the consistency equation which, if satisfied by a solution $\beta(z)$ of \eqref{nf5}, ensures that $\beta(z)$ corresponds to a solution of \eqref{nf2}.

Since $t_1(z) = \sqrt{1-t_2(z)^2-t_3(z)^2}=\sqrt{1-|\beta|^2}$, we have that
\be 
t_1 ' = - \frac{(|\beta|^2)'}{2\sqrt{1-|\beta|^2}} 
\ee
and 
\be 
t_2t_3' - t_2't_3 = \frac{\beta' \beta^* - \beta {\beta^*}'}{2i}\,,
\ee
where again $^*$ denotes complex conjugation. Yet, $\beta' \beta^* - \beta {\beta^*}' = 2i|\beta|^2(\text{arg}(\beta))'$, where $\text{arg}(\beta)$ denotes the complex argument of $\beta$. With this, equation \eqref{nf3a} can be put into the form
\be \label{cc1}
(1-\alpha')|\beta|^2 (\text{arg}(\beta))' - \frac{\alpha}{2}\frac{(|\beta|^2)'}{\sqrt{1-|\beta|^2}} - \alpha' U (1-|\beta|^2) + c\sqrt{1-|\beta|^2} =I_1\,.
\ee 
This is the consistency equation for $\beta(z)$. If $\beta(z)$ is a solution to the equation \eqref{nf5} and to the consistency equation \eqref{cc1}, then $\textbf{t}(z) = [\sqrt{1-|\beta(z)|^2},\text{Re}(\beta(z)),\text{Im}(\beta(z))]^T$ is a solution to \eqref{nf2}.

Often, this consistency equation is satisfied if the constant $I_1$ is properly selected. While $I_2$ and $I_3$ can usually be set to zero, the value of $I_1$ depends on the requirement for consistency of the solution $\beta(z)$. For example, let us revisit the helical solution \eqref{helixagain}. While this function satisfies the equation \eqref{nf5}, placing this form of $\beta(z)$ into the consistency equation \eqref{cc1}, we find
\be 
I_1 = c\sqrt{1-A^2} + U(A^2 - \alpha')\,.
\ee
Therefore, the choice of the arbitrary constant $I_1$ is fixed by this specific solution. Choosing a different form for $\beta(z)$ will result in the specification of a different value for $I_1$. Therefore, a helical solution exists for the first integral \eqref{nf2} governing the traveling wave solution $\mathbf{t}(z)$ provided that the constant vector $\mathbf{I}$ is of the form $\mathbf{I}=[c\sqrt{1-A^2} + U(A^2 - \alpha'), 0,0]^T$.

\section{Waves driven by the normal fluid: dynamics on a sphere}
The potential formulation used in the previous section is useful when the vortex filament is essentially aligned along a single axis (in the case considered there, aligned along the direction of the normal fluid flow). In such a case, the other two components of the tangent vector may exhibit some symmetry, so it makes sense to consider a potential function $\beta(z)$ in order to study the traveling waves. However, when there is no obvious symmetry (or when the specific solution does not call for such symmetry), one may want to consider a different formulation. 

Again assuming that the normal fluid velocity is oriented like $\mathbf{U}=[U,0,0]^T$, let us write the unit tangent vector in spherical coordinates (on a sphere of radius one, since $|\bt|=1$ by the unit vector assumption):
\be \label{ox1}
\bt(z) = \left[ \cos(\Omega(z)), \sin(\Omega(z))\cos(\Xi(z)), \sin(\Omega(z))\sin(\Xi(z)) \right]^T\,.
\ee
This form of the tangent vector makes clear the fact that the solution falls on a sphere. As we change $z$, we obtain a trajectory on the unit sphere. Making use of the assumption \eqref{ox1}, equation \eqref{nf2} reduces into a system of ordinary differential equations for $\Omega(z)$ and $\Xi(z)$. We obtain
\be \label{ox2a}
(1-\alpha')\sin^2(\Omega)\Xi' - \alpha \sin(\Omega)\Omega' - \alpha' U \cos^2(\Omega) + c \cos(\Omega) = I_1\,,
\ee
\be \begin{aligned}\label{ox2b}
-(1-\alpha')&(\sin(\Xi)\Omega' + \sin(\Omega)\cos(\Omega)\cos(\Xi)\Xi') + \alpha (\cos(\Omega)\cos(\Xi)\Omega' - \sin(\Omega)\sin(\Xi)\Xi') \\
& + \alpha U \sin(\Omega)\sin(\Xi) - \alpha' U \sin(\Omega)\cos(\Omega)\cos(\Xi) + c \sin(\Omega)\cos(\Xi) = I_2\,,
\end{aligned}\ee
\be \begin{aligned}\label{ox2c}
(1-\alpha')&(\cos(\Xi)\Omega' - \sin(\Omega)\cos(\Omega)\sin(\Xi)\Xi') + \alpha (\cos(\Omega)\sin(\Xi)\Omega' + \sin(\Omega)\cos(\Xi)\Xi') \\
& - \alpha U \sin(\Omega)\cos(\Xi) - \alpha' U \sin(\Omega)\cos(\Omega)\sin(\Xi) + c\sin(\Omega)\sin(\Xi) = I_3 \,.
\end{aligned}\ee

We first assume that both angular coordinates $\Omega(z)$ and $\Xi(z)$ in the representation \eqref{ox1} are not constant in the wave variable, $z$.
Let us take $\cos(\Xi)$\eqref{ox2b} + $\sin(\Xi)$\eqref{ox2c}; we find
\be \label{ox3}
(1-\alpha')\sin(\Omega)\cos(\Omega)\Xi' - \alpha \cos(\Omega)\Omega' + \alpha' U \sin(\Omega)\cos(\Omega) - c \sin(\Omega) = -I_2 \cos(\Xi) - I_3 \sin(\Xi)\,.
\ee
There are three equations \eqref{ox2a}-\eqref{ox2c} for two unknown functions. We can view \eqref{ox2a} as a consistency condition for the solution of the two equations \eqref{ox2b}-\eqref{ox2c}. Therefore, \eqref{ox2a} and \eqref{ox3} should be consistent. If we calculate $\sin(\Xi)$\eqref{ox3} - $\cos(\Xi)$\eqref{ox2a}, we find that the consistency requirement is
\be 
\alpha' U \cos(\Omega) - c = - I_1 \cos(\Omega) - I_2 \sin(\Omega)\cos(\Xi) - I_3 \sin(\Omega)\sin(\Xi)\,.
\ee
We therefore have consistency provided that $I_1 = -\alpha' U$, $I_2=I_3 =0$, and $c=0$. Since $c=0$, we have $z=s$ and therefore this solution is a stationary solution, corresponding to a type of standing wave. Under these conditions, \eqref{ox2a} is consistent with \eqref{ox2b}-\eqref{ox2c}. 

Let us next take $\cos(\Xi)$\eqref{ox2c} - $\sin(\Xi)$\eqref{ox2b}; we find
\be \label{ox4}
(1-\alpha')\Omega' + \alpha \sin(\Omega)\Xi' - \alpha U \sin(\Omega) =0\,.
\ee
From equations \eqref{ox3} and \eqref{ox4}, we obtain the system
\be \begin{aligned} 
(1-\alpha')\sin(\Omega)\Xi' - \alpha \Omega' + \alpha' U \sin(\Omega) & = 0\,,\\
\alpha \sin(\Omega)\Xi' + (1-\alpha')\Omega' - \alpha U \sin(\Omega) & = 0\,.
\end{aligned}\ee
Rearranging this system, we find that the equations decouple and we obtain
\be \label{ox5}
\Xi' = \frac{\alpha' -{\alpha'}^2 + \alpha^2}{(1-\alpha')^2 + \alpha^2}U \quad \text{and} \quad \Omega' = \frac{(1-2\alpha')\alpha U}{(1-\alpha')^2 + \alpha^2} \sin(\Omega)\,.
\ee
To obtain one particular closed form solution (for sake of example), we shall impose a set of initial conditions on this ODE system. Solving \eqref{ox5} subject to $\Xi(0)=0$ and $\Omega(0)=\frac{\pi}{2}$, we have
\be \label{8.10}
\Xi(s) = \frac{\alpha' -{\alpha'}^2 + \alpha^2}{(1-\alpha')^2 + \alpha^2}Us\,,
\ee
\be \label{8.11}
\Omega(s) = \text{arctan}\left(  \text{sech}\left( \frac{(1-2\alpha')\alpha U s}{(1-\alpha')^2 + \alpha^2} \right), -\tanh\left( \frac{(1-2\alpha')\alpha Us}{(1-\alpha')^2 + \alpha^2} \right)  \right)\,.
\ee
Here we use the two-argument arc-tangent function, which can be defined in terms of the standard arctangent function $\text{tan}^{-1}$ by the relation
\be 
\text{arctan}(a_1,a_2) = 2 \text{tan}^{-1}\left( \frac{a_1}{\sqrt{a_1^2 + a_2^2} - a_2} \right) \quad \text{for} \quad a_1 \neq 0\,.
\ee
When both angular coordinates are non-constant, we therefore obtain (from the consistency requirement) standing waves. Note that while the wave speed is zero from the consistency equation, the normal fluid velocity is still essential for generating these solutions. In the limit where $U\rightarrow 0$, the tangent vector reduces to a constant vector $\bt(s)=[0,1,0]^T$. A similar limit is reached when $\alpha,\alpha' \rightarrow 0$. We also remark that in this case, the constant vector $\mathbf{I}$ in \eqref{nf2} is not the zero vector. Rather, we have that $\mathbf{I}=[-\alpha' U, 0,0]^T$ in order to satisfy the consistency condition. In Fig. 6, we plot some of these solutions on the sphere (for reference).

\begin{figure}
\begin{center}
\includegraphics[scale=0.3]{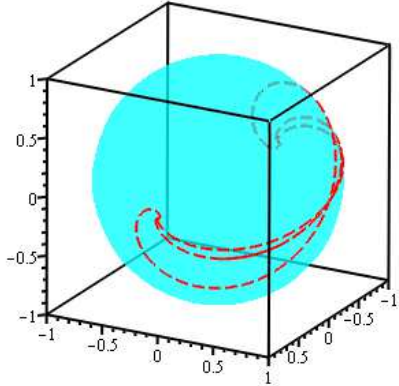}
\end{center}
\caption{(Color Online) Stationary tangent vector ($\mathbf{t}(z)=\mathbf{t}(s)$) dynamics on the unit sphere with \eqref{ox1} given the specific solutions \eqref{8.10} and \eqref{8.11}. For various combinations of the mutual friction parameters, we plot the exact solution over $s\in \mathbb{R}$. The $s\rightarrow -\infty$ and $s\rightarrow +\infty$ points are antipodal points on the sphere. The intermediate dynamics are controlled by the mutual friction terms. Since $U$ scales the variable $s$ directly, there are no qualitative effects from $U$ to the structures obtained if we plot $s\in \mathbb{R}$. (In contrast, the mutual friction parameters scale $s$ nonlinearly.) However, note that $U\neq 0$ is essential here, otherwise for $U=0$ we just obtain a fixed point on the sphere. This point corresponds with $s=0$, which is a fixed point of all of the maps (the curves all intersect when $s=0$, at the position $\bt(s=0) = [0,1,0]^T$). \label{Fig6}}
\end{figure}

When the angular coordinate $\Omega$ is constant in $z$, we can recover solutions which do depend on the wave speed $c$. Once such can be found by taking, say, $\Omega = \frac{\pi}{4}$, which results in a helical filament (as described through another approach in the previous section). Likewise, we can obtain distinct structures when taking $\Xi$ to be a constant in $z$. For instance, setting $\Xi=0$, we can construct a vortex filament ring.

\section{Waves driven by the normal fluid: direct consideration of the quantum LIA}
Let us now turn our attention directly to the vector LIA \eqref{fil} governing the motion of a filament curve given in vector form by a function $\mathbf{r}$, where $\mathbf{r}_t = \mathbf{v}$ and $\mathbf{r}_s = \bt$. Under certain nice conditions (namely, when $|\mathbf{U}|$ was negligibly small), we were able to convert \eqref{fil} into a scalar PDE by means of a quantum Hasimoto transformation (Section 2), and this transformation was used to study a number of solutions (Sections 3-6). For other cases, we were able to study the tangent vector $\bt(s,t)$ in the presence of non-negligible normal fluid velocity $\mathbf{U}$ (Sections 7-8). In all of these cases, we exploited various simplifications or symmetries to arrive at nice analytical results describing a variety of filament configurations through a study of the tangent vector to the filament (the position curve for the filament could then be reconstructed through an integration with respect to arclength). We shall now consider the quantum LIA \eqref{fil} directly in terms of the position vector of the filament (a curve $\mathbf{r}(x,t)$, where $x$ is some spatial parameter, not necessarily arclength), without such simplifications, since this is needed in situations where the various simplifications and symmetries exploited previously are no longer useful or are perhaps not quite so useful. Writing the filament curve as a function of time and a spatial parameter (say $x\in \mathbb{R}$), \eqref{fil} reads
\be \label{rvect}
\textbf{r}_t =  \frac{\mathbf{r}_x \times \mathbf{r}_{xx}}{|\mathbf{r}_x|^3} + \alpha \frac{\mathbf{r}_x}{|\mathbf{r}_x|} \times \left(\mathbf{U} - \frac{\mathbf{r}_x \times \mathbf{r}_{xx}}{|\mathbf{r}_x|^3}\right) - \alpha^\prime \frac{\mathbf{r}_x}{|\mathbf{r}_x|} \times \left(\frac{\mathbf{r}_x}{|\mathbf{r}_x|} \times \left(\mathbf{U} - \frac{\mathbf{r}_x \times \mathbf{r}_{xx}}{|\mathbf{r}_x|^3} \right)\right)\,.
\ee
Although the choice of a spatial parameter $x$ other than arclength complicates this formula slightly, this enables us to study the position vector $\mathbf{r}$ without worrying about the dependence of arclength $s$ on the magnitude $|\mathbf{r}_x|$. Unlike in the case of the tangent vector $\bt$ with $|\bt|=1$, we do not necessarily need to assume that $|\mathbf{r}_x|$ satisfies any scaling property. So, we obtain terms of the form $\bt = \mathbf{r}_s=\frac{\mathbf{r}_x}{|\mathbf{r}_x|}$. The formulation \eqref{rvect} has the benefit of giving us a solution in extrinsic (or, laboratory) coordinates, as opposed to the intrinsic arclength parameterization. There are multiple types of solutions one may construct in this frame (and that have already been constructed), so in keeping with the theme of the present paper, we shall limit our attention to waves which are driven by the normal fluid flow. Kelvin waves driven by the normal fluid flow were considered under the extrinsic coordinate framework by Van Gorder (2014b), so we shall exclude such solutions from consideration here.

\subsection{Structures maintaining their shape as they propagate with the normal fluid}
Let us consider a sort of separable solution, $\mathbf{r}(x,t)=\hat{\mathbf{r}}(x)+\mathbf{U}t$. Provided such a solution exists, it would maintain its spatial form (given by the first term) while moving along with the normal fluid velocity. We may show that such a solution always exists, provided that $\hat{\mathbf{r}}(x)$ is a solution to
\be \label{planar1}
\frac{\hat{\mathbf{r}}_x \times \hat{\mathbf{r}}_{xx}}{|\hat{\mathbf{r}}_x|^3} = \mathbf{U}\,.
\ee
Indeed, if this condition holds, the two friction terms in \eqref{rvect} vanish, and due to $\mathbf{r}_t = \mathbf{U}$, we exactly recover the constraint \eqref{planar1}. Therefore, if a solution to \eqref{planar1} exists, a solution $\mathbf{r}(x,t)=\hat{\mathbf{r}}(x)+\mathbf{U}t$ to \eqref{rvect} must exist. Such solutions have the interesting property that they are not influenced by the superfluid friction parameters, even though they are influenced by the normal fluid flow. Let us then consider solutions of equation \eqref{planar1}.

From geometry of space curves, recall that if the binormal vector for a space curve $\hat{\mathbf{r}}$ is constant, then that curve is planar. Here, we have the binormal vector with spatial scaling $x$ and a scaling factor $|\hat{\mathbf{r}}_x|^{-3}$, as we have the binormal vector times the curvature. This total quantity is equal to a constant vector. While this is not exactly the same as the binormal itself being constant, let us still assume a planar solution, as this will help us to demonstrate the existence of a solution $\hat{\mathbf{r}}(x)$ to \eqref{planar1}. Let us orient our coordinate system so that $\mathbf{U}=[U,0,0]^T$. Then, we seek a planar solution which is orthogonal to this normal fluid velocity, viz., $\hat{\mathbf{r}}(x)=\frac{1}{U}[0,f(x),g(x)]^T$, where $f(x)$ and $g(x)$ are twice-differentiable functions. The scaling of $\frac{1}{U}$ simplifies the resulting mathematics, and highlights the fact that the solution depends on a non-zero normal fluid velocity. Equation \eqref{planar1} then becomes
\be \label{planar2} 
\frac{f'g'' - f''g'}{({f'}^2+{g'}^2)^{3/2}} = 1\,,
\ee
where prime denotes differentiation with respect to $x$. We find that this equation is satisfies provided that the relation $f(x)^2 + g(x)^2 =1$ holds. 

For instance, if we set $f(x)=\cos(x)$ and $g(x)=\sin(x)$, we obtain the solution
\be 
\mathbf{r}(x,t) = \left[ Ut, \frac{\cos(x)}{U} , \frac{\sin(x)}{U} \right]^T \,.
\ee
This solution describes a vortex ring of radius $\frac{1}{|U|}$ oriented orthogonal to the direction of the normal fluid velocity. Since the ring is orthogonal to the normal fluid velocity, it is able to maintain its form and propagates exactly with the normal fluid. As the normal fluid velocity increases in magnitude, the rate of propagation increases, while the radius of the ring contracts. So, rapidly moving rings will have small radii, whereas slowly moving rings can have large radii. The ring is eternal, and continues to maintain its shape as it moves in the direction of the normal fluid velocity. If, instead, we were to obtain a ring not in a plane orthogonal to the normal fluid flow, we would find that the normal fluid velocity would be disruptive, and also that the mutual friction terms would cause decay of the vortex ring. Therefore, this ``eternal" ring can exist only with an ideal orientation relative to the normal fluid velocity. 

Another example of a solution would be found upon taking $f(x)=\tanh(x)$ and $g(x)=\text{sech}(x)$, which gives
\be 
\mathbf{r}(x,t) = \left[ Ut, \frac{\tanh(x)}{U} , \frac{\text{sech}(x)}{U} \right]^T \,.
\ee
Unlike the closed vortex ring, this describes an open filament which forms an arc. This appears as a topological soliton with the ``hump" of the arc due to the sech term (although more interesting topological solitons could be constructed with different functions $f(x)$ and $g(x)$). The ends of the arc correspond to the rapid decay as the parameter $x$ approaches $x\rightarrow \pm \infty$. Again, the filament moves exactly with the normal fluid velocity, and the size of the spatial structure scales inversely with the magnitude of the normal fluid velocity. 

Since the conditions on the functions $f(x)$ and $g(x)$ are fairly general (we essentially have one free function $f(x)$ satisfying $|f(x)|\leq 1$), we can construct any number of open or closed filaments with planar spatial structures. These filaments all are oriented orthogonal to the normal fluid velocity, move with the normal fluid, and are spatially scaled in size by the magnitude of the normal fluid velocity. Due to the orientation of such filaments, they do not feel the mutual friction effects, and hence maintain their spatial structure eternally (unless acted upon by an external force or boundary).

These planar structures are rather specialized, in that they lie orthogonal to the direction of the normal fluid velocity. This has the effect of shielding the solutions from the mutual friction effects, and the solutions are able to propagate without a change in form. In contrast, when initially planar solutions are not impacted on orthogonally by the normal fluid velocity, there will be torsion effects due to mutual friction which gradually deform the filaments. In such a case, an initial planar filament will gradually bend and rotate due to the mutual friction effects. This type of behavior was studied in detail recently by Van Gorder (2014c).

\subsection{Chaos from traveling wave solutions to the quantum LIA?}
Throughout this paper, we have considered a wide variety of solutions to the quantum LIA \eqref{fil}. While the form of these solutions has varies, they all display various symmetries and rather regular structures. Essentially, the dynamics considered have been one or two dimensional (dynamics arising from the quantum Hasimoto transform (Sections 2-6) or from direct consideration of the unit tangent vector (Sections 7-8) are at most two dimensional). In the example above, the eternal solution dynamics are one-dimensional, with the solutions moving exactly with the normal fluid velocity. However, for chaotic dynamics it is known that we must have at least a three-dimensional system.

Let us consider a traveling wave solution to vector partial differential equation \eqref{rvect}. To make computations more tractable, let us write $\mathbf{r}(x,t)=\mathbf{r}(z)=\int^z_0 \mathbf{w}(\overline{z})d\overline{z}$, where $z=x-ct$, $c\in \mathbb{R}$ is the wave speed, and we assume $\mathbf{w}$ is differentiable. Then, \eqref{rvect} is put into the form
\be \label{wvect}
-c\textbf{w}  =  \frac{\mathbf{w} \times \mathbf{w}'}{|\mathbf{w}|^3} + \alpha \frac{\mathbf{w}}{|\mathbf{w}|} \times \left(\mathbf{U} - \frac{\mathbf{w} \times \mathbf{w}'}{|\mathbf{w}|^3}\right) - \alpha^\prime \frac{\mathbf{w}}{|\mathbf{w}|} \times \left(\frac{\mathbf{w}}{|\mathbf{w}|} \times \left(\mathbf{U} - \frac{\mathbf{w} \times \mathbf{w}'}{|\mathbf{w}|^3} \right)\right)\,,
\ee
where prime denotes differentiation with respect to $z$. Using various vector simplifications, we obtain the first-order vector ordinary differential equation
\be 
(1-\alpha')\frac{\mathbf{w}\times\mathbf{w}'}{|\mathbf{w}|^3} + \alpha\left\lbrace \frac{\mathbf{w}'}{|\mathbf{w}|^2} - \frac{(\mathbf{w}\cdot\mathbf{w}')\mathbf{w}}{|\mathbf{w}|^4} \right\rbrace + \alpha \frac{\mathbf{w}\times\mathbf{U}}{|\mathbf{w}|} - \alpha' \frac{(\mathbf{w}\cdot\mathbf{U})\mathbf{w}}{|\mathbf{w}|^2} + \alpha' \mathbf{U} + c\mathbf{w} = \mathbf{0}\,.
\ee
This vector ordinary differential equation is equivalent to a system of three nonlinear and highly coupled ordinary differential equations. We may write this system in the matrix form
\be \label{ode}
\mathbf{M}(\mathbf{w})\mathbf{w}' = \mathbf{N}(\mathbf{w},\mathbf{U},c)\,,
\ee
where
\be \begin{aligned}
& \mathbf{M}(\mathbf{w}) =\\
& \left[\begin{matrix}
\alpha (w_2^2 + w_3^2) & - (1-\alpha')w_3|\mathbf{w}|-\alpha w_1w_2  &  (1-\alpha')w_2|\mathbf{w}| - \alpha w_1w_3 \\
(1-\alpha')w_3|\mathbf{w}| - \alpha w_1w_2 & \alpha(w_1^2+w_3^2) & - (1-\alpha')w_1|\mathbf{w}| -\alpha w_2w_3]\\
-(1-\alpha')w_2|\mathbf{w}| +\alpha w_1w_3] & (1-\alpha)w_1|\mathbf{w}| -\alpha w_2w_3 & \alpha(w_1^2+w_2^2)
\end{matrix}\right]\,,
\end{aligned}\ee
and 
\be 
\mathbf{N}(\mathbf{w},\mathbf{U},c) = \alpha |\mathbf{w}|^3 (\mathbf{U}\times\mathbf{w}) + \alpha' |\mathbf{w}|^2 (\mathbf{U}\cdot\mathbf{w})\mathbf{w} - \alpha' |\mathbf{w}|^4 \mathbf{U} - c|\mathbf{w}|^4\mathbf{w}  \,,
\ee 
where we write $\mathbf{w}(z)=[w_1(z),w_2(z),w_3(z)]^T$ and $|\mathbf{w}|=\sqrt{w_1^2+w_2^2+w_3^2}$. If one attempts to solve the system \eqref{ode} subject to any initial conditions, one will quickly find that an analytical or numerical solution is not forthcoming. This is not due to any complications from nonlinearity; rather, the system is degenerate for all $\mathbf{w}$. Indeed, we find that $\text{det}(\mathbf{M}(\mathbf{w})) \equiv 0$ for all $\mathbf{w}$, so the matrix $\mathbf{M}(\mathbf{w})$ is not full rank. What this means is that there is a dependency between the components of $\mathbf{w}$, hence we cannot solve for three unique functions $w_1(z)$, $w_2(z)$, $w_3(z)$ from \eqref{ode}. To make this system solvable, one would have to make assumptions on the relation between these components or one would have to specify a functional form for one of the functions (not unlike what has been done in earlier section of this paper). As such, the dynamics of \eqref{ode} are not fully three-dimensional, but rather are at most two-dimensional.

We can remark that there is a scaling of solutions which permits us to solve \eqref{ode}, under restrictive conditions. The homogeneous problem for the vector $\mathbf{w}'$, namely $\mathbf{M}(\mathbf{w})\mathbf{w}'=\mathbf{0}$ is identically satisfied whenever the derivatives of the three constituent functions satisfy the conditions
\be 
\frac{w_2'}{w_2} = \frac{w_1'}{w_1} = \frac{w_3'}{w_3}\,.
\ee
This implies that a solution of the form 
\be 
\mathbf{w}_{\text{particular}}(z)=[w_1(z),A_2w_1(z),A_3w_1(z)]^T=[1,A_2,A_3]^T w_1(z)
\ee
exists for the homogeneous problem $\mathbf{M}(\mathbf{w})\mathbf{w}'=\mathbf{0}$. This makes sense, as the terms in the matrix $\mathbf{M}(\mathbf{w})$ all arise from cross products of the type $\mathbf{w}\times\mathbf{w}'$, and such terms are identically zero for the particular solution $\mathbf{w}_{\text{particular}}(z)$. This is one example of when a reduction in the dimension of the equation can yield a solution to the system \eqref{ode}, even if the full system is degenerate. Of course, one would next have to determine the parameters and unknown function in the particular solution $\mathbf{w}_{\text{particular}}(z)$ through solving $\mathbf{N}(\mathbf{w}_{\text{particular}}(z),\mathbf{U},c)=\mathbf{0}$ in order to ensure that the solution is indeed consistent.

Now, since the dynamics governing a general traveling wave solution involve an added functional constraint $\text{det}(\mathbf{M}(\mathbf{w})) \equiv 0$, the dynamics of the system \eqref{ode} cannot give chaos. As the assumptions used to derive this equation were general, we can say that deterministic chaos is not expected from traveling wave solutions of full the quantum LIA. This is not to say that other structures might not yield deterministic chaos, but nonlinear traveling waves will not. However, due to inherent symmetries in the quantum LIA (which we exploit in order to obtain fairly regular dynamics for a variety of nonlinear waves on quantum vortex filaments), there is a propensity for the dynamics to be two-dimensional (essentially translational dynamics and rotational dynamics). Therefore, any more complicated dynamics would likely require very specific requirements on the vortex filament.

\section{Discussion}
We have successfully obtained a quantum Hasimoto transformation which maps the quantum LIA into a type of scalar partial differential equation \eqref{eveq} (essentially a complex cubic Ginzburg-Landau equation (GLE) with extra dissipative terms), by use of a method analogous to the Hasimoto transformation for a classical fluid vortex filament. We therefore refer to this procedure as the quantum Hasimoto trasnformation. Doing so, we are able to reduce the quantum LIA \eqref{unitLIA} (a vector conservation law) into a complex scalar PDE \eqref{eveq}, which makes the system far more amenable to mathematical analysis. Such a mapping between the quantum LIA and this PDE is also desirable from a physical point of view, since it allows for greater qualitative comparison of the quantum and classical fluid LIA solutions. 

Upon transforming the quantum LIA into a complex scalar PDE \eqref{eveq}, we were able to study a number of solutions. First we obtained Stokes wave type solutions in Section 3. In the case of a classical fluid, these solutions take the form of oscillating waves with constant amplitudes. However, we were able to demonstrate that for a quantum fluid modeled under LIA, such solutions have an algebraic decay rate and therefore dissipate as time becomes large. The period of oscillation for such solutions is variable, as well, and gradually increases in time. Since the function $\psi$ used in this paper is a composite function of curvature and torsion, the physical interpretation for these solutions to the quantum LIA is that the curvature of the filament decreases in time, in contrast to the classical fluid solutions, where curvature is constant. In both the classical and quantum setting, the torsion for these Stokes waves is zero. The result is not surprising in light of the fact that the complex scalar PDE is dissipative and hence one would expect this type of solution to exhibit such decay. The rate of decay for these waves is algebraic in time, of the order $t^{-1/2}$. Since curvature of the waves along a quantum filament will be maximized a small timescales, and gradually decays for large time, this means that any waves or disturbances of this type are expected to be short lived. Physically, any such perturbations along quantum vortex filaments will have a tendency to be smoothed out as time progresses, which makes sense in light of the dissipation. Note that under the present framework, we do not have a Kelvin waves driven by the normal fluid flow, as we take $\mathbf{U}=O(\alpha)$, and hence these contributions become negligible. Such Kelvin waves were studied analytically under the model \eqref{fil} in Van Gorder (2014b) for both constant amplitude and time-variable amplitude waves under a different framework which allows for the inclusion of larger order effects from $\mathbf{U}$. Likewise, there is no purely planar filament corresponding to a solution under the present framework. The time evolution of a quasi-planar filament can, however, be described for the model \eqref{fil}. Such a filament starts out planar at $t=0$ and gradually deforms due to torsion effects coming from the friction terms. Properties of this solution were discussed in Van Gorder (2014c).

A second and perhaps more fundamental solution is that of the soliton on a vortex filament. Hasimoto originally employed the aforementioned transform in order to demonstrate the existence of a soliton on a vortex filament under the classical LIA. In Section 4, we have been able to demonstrate analogously that a standing soliton also exists under the quantum LIA. The soliton takes the form of a sech function (i.e., a bright soliton), which is what one finds for the classical fluid case, plus a perturbation of order $\alpha$ which involves time. The inclusion of mutual friction parameters results in the appearance of an additional phase factor that depends on arclength and time. This means that the torsion effects on the classical standing wave are enhanced, as one might expect since the wave must contend with friction under the quantum LIA. The curvature maxima persists up to $O(\alpha)$, meaning that any dissipative effects on the curvature are confined to order $O(\alpha^2)$ terms, which are very small for temperatures below 2.0K. Therefore, while the soliton is not eternal (as is true in the classical case), the soliton would dissipate at a rate proportional to at most $\alpha^2$, which is rather slow. As such, the perturbation of the standing sech wave should persist over a reasonably large time interval, and this in turn implies that such waves should be obtainable from low-temperature experiments. Note that the Hasimoto formulation has proven useful in experiments (Hopfinger and Browand, 1982), and we expect the present results should be similarly useful for experiments in superfluid vortex dynamics, due to the persistence of this standing wave solution. We should also remark that breather solitons have been found for the classical LIA (Umeki, 2010; Salman, 2013) (with no mutual friction parameters present), and one extension of the present paper would be to consider breather solitons for the quantum LIA. It may be the case that the amplitudes (curvature) of these breathers will remain more-or-less the same, with only strong phase modifications present due to the inclusion of mutual friction terms (and the resulting increase in torsion). 

Self-similar solutions were constructed in Section 5 under the quantum Hasimoto map, and these solutions demonstrate that local regions of increased torsion can develop for positive time. However, as time becomes asymptotically large, these regions are smoothed and the torsion and curvature tend toward zero. Thus, finite kinks and sharp turns along the quantum vortex filament are gradually smoothed by the friction effects. The results we obtain allow one to recover the results of Lipniacki (2003a,b), since by obtaining a self-similar solution for the scalar equation \eqref{eveq}, once is able to recover the curvature and torsion found directly by Lipniacki through a direct approach (in those works, curvature and torsion were kept two separate quantities, and were solved as a system).

The final solutions obtained using the quantum Hasimoto transformation were traveling wave solutions which propagate along the quantum vortex filament with constant wave speed $c$. Such solutions, as discussed in Section 6, generalize the standing soliton of Section 4. In addition to sech-type waves, waves with oscillating periodic curvature are also possible (in analogue to the elliptic sn or cn wave solutions possible in the cubic NLS equation), and we demonstrate these waves numerically. It makes sense to consider small magnitude wave speeds. For large wave speeds, the solutions appear to lose stability, and for very large wave speeds, the solutions may fail to exist (although we have not shown this rigorously). Note that in the case where $\mathbf{U}$ is large, the normal fluid flow allows for the propagation of Kelvin waves along a filament, with wave speed a function of $|\mathbf{U}|$. The same is likely true of these nonlinear waves, and an area of future work will involve the study of such nonlinear wave solutions to \eqref{fil} when $\mathbf{U}$ is large relative to other terms in the model. To study such waves, we can no longer use the quantum Hasimoto transform (valid when $|\mathbf{U}|$ is small enough so as to be negligible), as one will need to include the strong effects of $\mathbf{U}$. 

Derivations using the quantum Hasimoto transformation exclude any strong effects from the normal fluid velocity vector, $\mathbf{U}$. It is possible to include the effects of the normal fluid, although although one cannot use the scalar PDE derived under the Hasimoto transformation. This is due to the fact that for a general vector $\mathbf{U}$ there is no Hasimoto mapping from the quantum LIA \eqref{unitLIA} to a scalar equation like \eqref{eveq}, and one must instead attempt to solve for the tangent vector function $\mathbf{t}$ directly or through some other reduction. This is what we have done in the remainder of the paper (Sections 7-9). In Section 7, we derive a scalar ODE governing the tangent vector dynamics under the assumption of a traveling wave solution. Some exact solutions are given to illustrate the approach. This representation is valid when the waves take on a traveling wave form given some fixed wave speed, $c$. There is an additional consistency equation (which essentially ensures that the tangent vector solution remains a unit vector) that must be satisfied under this formulation, and we give the general form of this condition. A concrete example is given to demonstrate the manner in which this consistency requirement is met for a specific class of traveling wave solutions.

In Section 8, we map the tangent vector dynamics into a dynamical system on the unit sphere. The resulting dynamical system is then naturally two-dimensional (as the sphere is a two-dimensional manifold). We give a class of solutions to this system as an example of such dynamics, which can then be integrated in order to recover the position vector from the tangent vector. These dynamics are also plotted on the sphere for various values of the model parameters (see Fig. 6), so as to illustrate the results. Other dynamics arising from these equations can be used to construct filament structures such as helical vortex filaments and vortex filament rings.

In Section 9 we directly consider the quantum LIA for the position vector $\mathbf{r}(x,t)$, where $x$ is some parameterization corresponding to a laboratory length coordinate. We first obtain a class of solutions which are oriented in a plane orthogonal to the direction of the normal fluid velocity. These filament structures move along with the normal fluid velocity while maintaining their form, as they are oriented in such a way that allows them to escape the normal and binormal mutual friction effects. While such panr solutions are rather specialized, a number of specific solution forms can be constructing. One particularly interesting solution is the vortex filament ring. The plane of the ring is orthogonal to the direction of the normal fluid velocity and propagates exactly with the normal fluid. The radius of the ring is determined by the magnitude of the normal fluid velocity, with the solution tightening up as the magnitude of the normal fluid velocity increases. This, and the other solutions oriented in this way, are eternal, in that they maintain their form as the propagate.

Also in Section 9, we derive the full quantum LIA for traveling wave solutions (depending again on a fixed wave speed $c$) to the vector PDE governing the position vector $\mathbf{r}(x,t)$ in the laboratory coordinates (that is, we consider $\mathbf{r}(x,t)=\mathbf{r}(x-ct$. The resulting vector ODE can be put into a system of three strongly coupled ODEs for the scalar components of $\mathbf{r}(x-ct)$. However, we find that this system is degenerate for any choice of arbitrary vector $\mathbf{r}(x-ct)$ (it is not full rank), and hence the dynamics cannot be fully three-dimensional in the case of such traveling waves. What this suggests, then, is that the two-dimensional dynamics considered elsewhere in the paper will always be sufficient to describe the traveling wave solutions. Importantly, this tells us not to expect chaotic dynamics from the nonlinear traveling waves appearing along a quantum vortex filament under the quantum form of the LIA, as chaotic dynamics are only possible for continuous dynamical systems of order three or greater. This rules out traveling waves as a possible route to chaos under the quantum LIA (or any of its reductions).

While we did not find complicated dynamics such as chaos for the traveling nonlinear wave solutions considered, there are other possible solutions to the PDE \eqref{eveq}, so more complex dynamics are certainly possible, perhaps for more exotic vortex filament structures. However, the structure of the filaments would have to counterbalance the dissipation effects of the mutual friction. This could be possible, under the right normal fluid velocity (which we have seen above enables eternal solutions for some special case). Without the normal fluid velocity, the solutions should dissipate at some timescale (even if this is a slow timescale relative to the timescale on which the solution propagates). Therefore, solutions to the PDE resulting from the quantum Hasimoto transform are likely not chaotic.  However, chaos has been shown to arise from related models (Aranson and Kramer, 2002), so perhaps there is a way ti mitigate the dissipative nonlinearity in the equation. Additionally, Lipniacki (2003b) considered self-similar solutions for the quantum LIA (not in the scalar model \eqref{eveq}, but directly for curvature and torsion) and hypothesized that chaotic solutions could be possible. Preliminary work suggests that the sign of the function $A(t)$ in the formulation developed in \eqref{eveq} could influence the bifurcation structure of traveling wave solutions, but as we mentioned above, these bifurcations will not lead to chaos (for the traveling wave case). 

We should note that chaotic dynamics of vortex filament solutions to the LIA were studied under the ``direct" interaction approximation (Nemirovskii and Baltsevich, 2001), in the stochastic sense. Therefore, there could be stochastic formulations for the quantum LIA which would permit chaos in this manner. (Although, due to the stochastic contribution, this would no longer be the classical ``deterministic" chaos.) Another possibility for chaos would be the interaction of multiple quantum vortex filaments, including collisions and reconnection events. Such dynamics have been considered numerically, but due to the complexity of the problem, analytical results are wanting. To be accurate in the evaluation of these dynamics, non-locality would also likely have to be considered. Recall that the classical LIA approximates the Biot-Savart dynamics governing the self-induces motion of a vortex filament. Analogously, the quantum LIA can be cast in a corresponding non-local form while maintaining the normal fluid and mutual friction effects. 

Here we have studied nonlinear waves along a single quantum vortex filament. These filament configurations constitute one part of the possible dynamics of the vortex lines. Superfluid turbulence involves large deformations of vortex loops, collisions and reconnection of vortex lines, the escape of vortex loops from the bulk, and so on. Kelvin waves have been considered as mechanisms for carrying energy to small scales, but this assignment is complicated (Nemirovskii, 2013). To account for such behaviors, one typically needs the non-local formulation of the vortex filament problem. Obtaining a scalar relation such as \eqref{eveq} in any general way for such a problem has not yet been done, although it may be possible for some specific vortex filament configurations to write down the relevant integral equation governing curvature and torsion.

Aside from more complicated dynamics, one area of interest would be a study of the stability of the solutions obtained under the quantum LIA. While a variety of solutions have been found, some even eternal (in that they maintain their form over time), it is natural to wonder what will happen to such solutions under small perturbations of their structure. For the classical LIA, some stability results of this kind are reported for standard vortex filament configurations (Banica \& Vega, 2009; Calini \& Ivey, 2011; Kida, 1982; Tsai \& Widnall, 1976; Van Gorder, 2013b; Widnall, 1972; Widnall \& Tsai, 1977). Extending such results to the quantum LIA would involve not only perturbations to the vortex filament solutions, but also structural perturbations to the governing equation which are necessary to account for the terms giving mutual friction effects. Some quantum results have been given for specific flow structures (Godfrey \& Samuels, 2000).

\end{document}